\documentclass[12pt,aps,prd,floatfix,nofootinbib,superscriptaddress,a4paper]{revtex4-2}
\pdfoutput=1

\usepackage{amsmath, amssymb, amsfonts, amsthm, latexsym, epsfig, mathrsfs, xcolor, bbm, slashed, braket}

\usepackage[all]{xy}

\usepackage[inline]{enumitem}

\usepackage{setspace}
\usepackage[marginal, multiple]{footmisc}

\usepackage[T1]{fontenc}
\usepackage[utf8]{inputenc}
\usepackage{lmodern}

\usepackage[colorlinks, allcolors=blue!70!black, linktocpage]{hyperref}

\definecolor{indigo(dye)}{rgb}{0.0, 0.25, 0.42}

\numberwithin{equation}{section}

\usepackage{cleveref}

\usepackage{microtype}

\usepackage{floatrow}
\let\OLDtableofcontents\tableofcontents
\renewcommand\tableofcontents[1]{%
    {\baselineskip 0.5ex %
	\OLDtableofcontents{#1}}%
}

\let\OLDthebibliography\thebibliography
\renewcommand\thebibliography[1]{%
	\setstretch{1.079} 
	\OLDthebibliography{#1}%
	\small %
	\setlength{\itemsep}{0.2\baselineskip} 
}

\let\OLDfootnote\footnote
\renewcommand\footnote[1]{%
	\setlength{\footnotesep}{0.75\baselineskip}%
	{\footnotesize \OLDfootnote{#1}}%
}

\setlist[enumerate]{noitemsep, label=(\arabic*), ref=(\arabic*)}

\newlist{condlist}{enumerate}{2}
\setlist[condlist,1]{itemsep=0.5pt, topsep=0pt, label=(\arabic*), ref=(\arabic*)}
\setlist[condlist,2]{noitemsep, label=(\alph*), ref=(\arabic{condlisti}.\alph*)}
\crefname{condlisti}{condition}{conditions}
\crefname{condlistii}{condition}{conditions}

\newlist{propertylist}{enumerate}{1}
\setlist[propertylist,1]{noitemsep, topsep=0pt, label=(\arabic*), ref=(\arabic*)}
\crefname{propertylisti}{Property}{Properties}

\renewcommand\thesection{\arabic{section}}
\renewcommand\thesubsection{\arabic{subsection}}

\makeatletter
\def\p@subsection{\thesection.}
\def\p@subsubsection{\thesection.\thesubsection.}
\makeatother 

\theoremstyle{plain}

\newtheorem{lemma}{Lemma}[section]

\theoremstyle{definition}

\theoremstyle{remark}
\newtheorem{remark}{Remark}[section]

\creflabelformat{equation}{#2#1#3}

\crefname{section}{\S}{\S}
\crefname{appendix}{Appendix}{Appendices}
\crefname{figure}{Fig.}{Figs.}
\crefname{table}{Table}{Tables}

\crefname{definition}{Def.}{Defs.}
\crefname{prop}{Prop.}{Props.}
\crefname{lemma}{Lemma}{Lemmas}
\crefname{corollary}{Cor.}{Cors.}
\crefname{thm}{Theorem}{Theorems}
\crefname{remark}{Remark}{Remarks}

\crefname{ass}{Assumptions}{Assumptions}
\crefname{property}{Properties}{Properties}

\newcommand{\be}{\begin{equation}\begin{aligned}}
\newcommand{\ee}{\end{aligned}\end{equation}}

\newcommand{\lb}{\left}
\newcommand{\rb}{\right}

\newcommand{\mc}{\mathcal}

\newcommand{\ms}{\mathscr}
\newcommand{\mf}{\mathfrak}
\newcommand{\bb}{\mathbb}

\newcommand{\eqsp}{\, ,\quad} 

\newcommand{\Lie}{\pounds} 
\newcommand{\defn}{\mathrel{\mathop:}=} 

\newcommand{\union}{\cup} 

\newcommand{\norm}[1]{\lb\Vert\, #1 \,\rb\Vert}		

\newcommand{\op}[1]{\boldsymbol{#1}}
\renewcommand{\1}{\op{1}}

\let\oldsetminus\setminus
\renewcommand{\setminus}{\!\oldsetminus\!} 

\newcommand{\pd}[2][]{\frac{\partial #1}{\partial #2}} 

\newcommand{\td}[2][]{\frac{d #1}{d #2}} 

\let\oldint\int
\renewcommand{\int}{\oldint\limits}

\let\oldlim\lim
\renewcommand{\lim}{\oldlim\limits}

\renewcommand{\bar}{\overline}

\newcommand{\grp}[1]{{\tt #1}}

\newcommand{\scri}{\ms I}

\newcommand{\Hilb}{\mathscr{H}}

\newcommand{\antiHilb}{%
\hspace{4pt} 
  \vbox{%
    \hrule height 0.5pt
    \kern0.25ex
    \hbox{%
      \kern-0.3em
      \ifmmode\Hilb\else\ensuremath{\Hilb}\fi
      \kern0em
    }
  }
}

\newcommand{\Fock}{\mathscr{F}}

\newcommand{\EM}{\textrm{EM}}

\newcommand{\YM}{\textrm{YM}}

\newcommand{\GR}{\textrm{GR}}

\newcommand{\nfrac}[2]{{{}^#1\!\!/\!_#2}}

\newcommand{\half}{\nfrac{1}{2}}

\newcommand{\orbit}{\ms{O}}
\newcommand{\oLie}{\mf{L}}

\newcommand{\abs}[1]{\lb\vert\, #1 \,\rb\vert}		

\newcommand{\s}{\omega}

\begin{document}

\setstretch{1.2}


\title{Infrared finite scattering theory: Scattering states and representations of the BMS group}

\author{Kartik Prabhu}
\email{kartikprabhu@rri.res.in}
\affiliation{Raman Research Institute, Sadashivanagar, Bengaluru 560080, India.}

\author{Gautam Satishchandran}
\email{gautam.satish@princeton.edu}
\affiliation{Princeton Gravity Initiative, Princeton University, Jadwin Hall, Washington Road, Princeton NJ 08544, USA}

\begin{abstract}
Any non-trivial scattering with massless fields in four spacetime dimensions will generically produce an ``out'' state with memory which gives rise to infrared divergences in the standard $S$-matrix. To obtain an infrared-finite scattering theory, one must suitably include states with memory. However, except in the case of QED with massive charged particles, asymptotic states with memory that have finite energy and angular momentum have not been constructed for more general theories (e.g. massless QED, Yang-Mills and quantum gravity). To this end, we construct direct-integral representations over the ``Lorentz orbit'' of a given memory and classify all ``orbit space representations'' that have well-defined energy and angular momentum. We thereby provide an explicit construction of a large supply of physical states with memory as well as the explicit action of the BMS charges all states. The construction of such states is a key step toward the formulation of an infrared-finite scattering theory. While we primarily focus on the quantum gravitational case, we outline how the methods presented in this paper can be applied to obtain representations of the Poincaré group with memory for more general quantum field theories.
\end{abstract}

\maketitle
\newpage 
\tableofcontents

\section{Introduction}
\label{sec:intro}

By far one of the most ubiquitous and successful formulations of quantum field theory is scattering theory in which, given a suitable collection of ``in'' states at past infinity, one obtains a map to a suitable collection of ``out'' states. At asymptotically early and late times, the asymptotic states are usually assumed to lie in the standard Fock space and the $S$-matrix is then defined as a map between the ``in'' and ``out'' standard Fock spaces. For massive fields, this map is well-defined  \cite{PhysRev.112.669,Ruelle1962,LSZ} but such an $S$-matrix generically encounters severe {\em infrared divergences} for theories with interacting, massless fields (e.g., QED or quantum gravity). This is because the massless field will generically undergo a {\em memory effect} where the leading order field (at $O(1/r)$ along null directions) at late retarded times (i.e. $u\to +\infty$) does not return back to its original value at early retarded times $(u\to - \infty)$. For example, in general relativity, the radiative field is the shear tensor $\sigma_{AB}$ at null infinity where the capital indices refer to angular indices on the sphere. The time-derivative yields the Bondi news tensor $N_{AB}=2\partial_{u}\sigma_{AB}$ and the memory $\Delta_{AB}(x^{C})$, in the gravitational case, captures the ``low-frequency'' behavior of the radiative field 
\begin{equation}
\label{eq:mem}
\Delta_{AB}(x^{C})=\frac{1}{2}\int_{\bb{R}}du~N_{AB}(u,x^{C})=\sigma_{AB}(u=+\infty,x^{C}) - \sigma_{AB}(u=-\infty,x^{C}) 
\end{equation}
where $x^{A}$ are angular coordinates on the sphere. In classical scattering, even if one starts with initial data with vanishing memory, the memory of the outgoing radiative field will be generically non-vanishing \cite{Zeldovich:1974gvh,PhysRevLett.67.1486,Wiseman:1991ss,Bieri_2013EM}. In the quantum theory, a non-vanishing memory implies that the expected number of radiative quanta diverges as $1/\omega$ at low frequencies and, consequently, the standard Fock space norm of states with memory similarly diverges \cite{narain81,Ashtekar:1981,asymp-quant}. 

In the context of collider physics --- where the relevant scattering timescales and detector resolution are insensitive to low-frequency physics --- one can impose an infrared cut-off and restrict attention to ``inclusive cross-sections'' for the scattering of any ``hard'' particles \cite{Bloch_1937,Yennie:1961ad,Weinberg:1965,Hannesdottir:2019umk}. Nevertheless, on longer time scales, the production of soft quanta cannot be ignored resulting in, for example, the complete decoherence\footnote{In four-dimensional, flat spacetime, this decoherence rate due to soft quanta grows logarithmically in time \cite{Carney_2017,Semenoff_2019}. However, in the presence of a black hole or cosmological horizon in {\em any} spacetime dimension, the analogous production of (massive or massless) soft quanta through the horizon results in a decoherence that grows linearly in time \cite{Danielson:2022tdw,Danielson:2022sga,Gralla:2023oya,Danielson:2024yru}.} of any ``hard'' particles at late times  \cite{Carney_2017,Semenoff:2018}. Furthermore, scattering theory as well as the asymptotically early and late time behavior of states play a central role in any holographic description of quantum gravity such as in celestial holography (see, e.g., \cite{Strominger:2017zoo,Pasterski:2021raf,Donnay:2023mrd,Pasterski:2023ikd} and references therein). However, since scattering will generically produce memory, the standard $S$-matrix cannot be defined since there are no states with memory in the standard Fock space. Therefore, in order to keep track of all observables --- including low-frequency observables such as the memory --- it is essential to formulate a well-defined, {\em infrared-finite} scattering theory. 

To obtain such an infrared-finite scattering theory one must enlarge the space of ``in'' and ``out'' states to include a sufficiently large number of states with memory. As we have mentioned, the standard Fock space $\Fock_{0}$ does not contain any states with memory and, as reviewed in \cref{sec:memFock}, states with memory $\Delta_{AB}$ lie in a different Fock space $\Fock_{\Delta}$ which is unitarily inequivalent to the standard Fock space \cite{Ashtekar:1981,asymp-quant}. As we will review shortly, while the states in $\Fock_{\Delta}$ have non-vanishing memory they do not, apriori, have well-defined energy-momentum and angular momentum. Furthermore, since these representations are labeled by the tensor $\Delta_{AB}$, there are uncountably many such inequivalent Fock representations and so they cannot all be contained in a single, separable Hilbert space. Additionally, the memory is not conserved and so any arbitrary separable subspace of states will not scatter into itself under evolution.

In QED --- where the analogous memory Fock representations $\Fock^{\EM}_{\Delta}$ are labeled by the electromagnetic memory $\Delta_{A}(x^{B})$ --- Faddeev and Kulish successfully assembled these Fock spaces together with states of the charged field to obtain a single, separable Hilbert space of {\em physical}, asymptotic states with well-defined energy-momentum and angular momentum which ``scatters into itself'' under evolution \cite{Kulish:1970ut,Chen_2009_I,Chen_2009_II}. Therefore, one can define an $S$-matrix on this ``Faddeev-Kulish'' Hilbert space \cite{Gabai:2016kuf,Duch:2019wpf}. The construction is based upon the charges associated with the generators of the infinite-dimensional, asymptotic symmetry group, i.e., the group of ``large gauge transformations'' at infinity. These charges are conserved \cite{KP-EM-match} and the Faddeev-Kulish Hilbert space corresponds to eigenstates of these conserved charges \cite{Kapec:2017tkm,narain81}. While the Faddeev-Kulish Hilbert space admits (a dense set of) physical states in QED\footnote{The ``large gauge charges'' are not Lorentz invariant and so the only Faddeev-Kulish Hilbert space which has a well-defined, strongly continuous action of the Lorentz group is the Hilbert space with vanishing large gauge charge \cite{Frohlich:1979uu}.}, it was proven in \cite{PSW-IR} that this construction cannot be generalized to QED with massless charged fields, Yang-Mills theories as well as quantum gravity and does not yield a large enough class of physical states. Indeed, in the gravitational case, the asymptotic symmetry group is the infinite-dimensional Bondi-van der Burg-Metzner-Sachs (BMS) group whose corresponding charges are conserved (see \cite{CE,Henneaux:2018gfi,Kartik_Maxwell,KP-GR-match,Mohamed:2021rfg,Mohamed:2023jwv}) and consist of --- in addition to the ordinary Poincaré charges --- an infinite number of supertranslation charges \cite{PhysRev.128.2851}. The analog of the Faddeev-Kulish Hilbert space corresponds to constructing eigenstates of the supertranslation charges, however, it was proven in \cite{PSW-IR} that the {\em only} physical eigenstate is the vacuum state.\footnote{While one can construct non-trivial eigenstates of the supertranslation charge in {\em linearized} quantum gravity \cite{Choi:2017bna,Choi:2017ylo}, these states do not have a well-defined action of the Lorentz group and therefore do not have a well-defined angular momentum (see section 6.3 of \cite{PSW-IR}).} Consequently, there does not appear to be any preferred Hilbert space for scattering in general quantum field theories and in quantum gravity. Nevertheless, it was argued in \cite{PSW-IR} that, in the absence of a preferred Hilbert space, one can still formulate a manifestly infrared-finite scattering theory where the asymptotic states are specified algebraically by their correlation functions on the in/out algebra of asymptotic operators. The scattering could then be computed, at least to any order in perturbation theory, using the Heisenberg evolution of operators. This formulation of scattering theory does not presuppose which Hilbert space the out state lies in and so is manifestly infrared finite. 

There is, however, a major issue in carrying out this proposal --- or any other proposed formulation of an infrared-finite scattering theory. As we have just explained, one does not require a ``preferred'' Hilbert space to formulate a well-defined, infrared-finite scattering theory. However, one does need to construct ``sufficiently many'' physical states with finite energy and angular momentum with non-vanishing memory. Given any in-state with finite energy and angular momentum, the out-state will generically be a state with finite energy, angular momentum and non-zero memory. While in QED with massive charges, the Faddeev-Kulish construction provides us with a large class of physical states with memory, states in massless QED, Yang-Mills or quantum gravity with finite energy-momentum, angular momentum and non-vanishing memory have never been constructed. 

This problem is a practical one and is well-illustrated by considering the example of the scattering of two gravitons. In this case, the ``in'' state is a state in the standard Fock space with finite energy and angular momentum. However, the ``out'' state will consist of outgoing ``hard'' gravitons together with soft radiation with memory. The precise distribution of memory depends strongly on the angular distribution of the incoming and outgoing gravitons.\footnote{This point is more precisely explained in \cite{KP-BMS-particles}.} For different incoming scattering angles, the ``out'' state will generally have different distributions in memory and therefore will lie in inequivalent Hilbert space representations.  The purpose of this paper is to construct a large supply of physical asymptotic states with finite energy, angular momentum and memory. We will do so by constructing an uncountably infinite number of unitarily inequivalent Hilbert spaces with memory that have a well-defined action of the BMS group. As we have already emphasized, the scattering theory cannot be formulated with a preferred (separable) Hilbert space and so we do {\em not} advocate attempting to construct an $S$-matrix using these representations. The scattering theory can be formulated in the absence of a preferred Hilbert space and the purpose is to characterize a large class of states which arise in scattering. 

For definiteness, we will focus primarily on the quantum gravitational case, however the construction proceeds in an analogous manner for other quantum field theories (see remark \ref{rem:qed-scalar-case}). We now give a general overview of the construction. As mentioned above, the Fock spaces $\Fock_{\Delta}$ yield a large class of states with definite memory $\Delta_{AB}$, however they do not correspond to {\em physical} states with well-defined energy and angular momentum. The energy of the gravitational waves is proportional to the integral over null infinity of the square of the Bondi news tensor $N_{AB}$ (i.e. the flux of energy is $\mathcal{F} = (-1/8\pi)\int_{\scri}dud\Omega~N^{AB}N_{AB}$ where $d\Omega$ is the measure on the $2$-sphere).
Therefore, by \cref{eq:mem}, a necessary condition for $\Fock_{\Delta}$ to admit a dense set of states with finite energy is that the memory $\Delta_{AB}(x^{A})$ must be square-integrable on the $2$-sphere. Thus, we restrict to those memory representations $\Fock_{\Delta}$ such that 
\begin{equation}
\label{eq:memL2}
\int_{\mathbb{S}^{2}}d\Omega~\Delta^{AB}\Delta_{AB} < \infty 
\end{equation}
is finite.\footnote{While it is tempting to restrict to smooth memories, instead of the bigger space of square-integrable ones, it is shown in \cite{KP-BMS-particles} that a generic scattering process necessarily produces non-smooth but square-integrable memory.} A much more serious issue arises when one seeks a well-defined action of the angular momentum. The memory is not invariant under general Lorentz transformations and, therefore, any eigenstates of memory in $\Fock_{\Delta}$ do not have well-defined Lorentz charges (see also \cite{asymp-quant}). Nevertheless, the space of memories obtained by acting on a given memory $\Delta_{AB}$ with the full Lorentz group is a finite-dimensional manifold --- the {\em orbit space} $\ms{O}$ associated to the given memory $\Delta_{AB}$. If $y\in \ms{O}$ denotes a point in the orbit space, the action of the Lorentz group yields a collection Fock spaces $\Fock_{\Delta(y)}$ over each point in $\ms{O}$. Furthermore, each orbit space admits a unique, Lorentz invariant measure $\mu(y)$. Thus, associated to each memory $\Delta_{AB}$, we obtain a {\em direct integral} Hilbert space\footnote{The direct integral given by \cref{eq:DIDelta} is over a finite-dimensional subspace of memories. One could instead attempt to consider a direct integral over the full, infinite-dimensional space of memories. This case was considered in \cite{Herdegen_1997} and \S~7 of \cite{PSW-IR} and requires a choice of infinite dimensional measure. While this choice is not unique, it was argued in \cite{PSW-IR} that there does not exist any Gaussian measure over the space of memories which yields a representation with well-defined energy and angular momentum.}
\begin{equation}
\label{eq:DIDelta}
\Fock_{\textrm{DI},\Delta} = \int_{\orbit}^{\oplus}d\mu(y)\Fock_{\Delta(y)}
\end{equation}
where the subscript ``$\textrm{DI}$'' refers to a direct integral. The precise definition of this direct integral is given in \cref{sec:Lorentzorb}. States on this direct integral Hilbert space do not have a definite memory, but have continuous distributions in memory over the orbit space $\orbit$. We show that, over the space of square-integrable memories, these Hilbert spaces carry a unitary representation of the full asymptotic BMS symmetry group and contain a dense set of states with non-zero memory, and well-defined energy and angular momentum. In particular, while the representation of the BMS supertranslations is straightforward, the action of the Lorentz group is highly non-trivial due to the fact that an infinitesimal Lorentz transformation changes the memory and thereby ``mixes'' the memory Fock spaces in the direct integral. Nevertheless, in \cref{sec:Lorentzorb}, we obtain explicit expressions for the unitary action of the Lorentz transformations and the Lorentz charges for a dense subspace of states in the direct integral Hilbert space. In \cref{sec:memory-orbits}, we adapt the analysis of McCarthy \cite{McCarthy2} to provide the classification of the orbit spaces of square-integrable memories that arise in gravitational scattering as well as their Lorentz invariant measures. We focus primarily on the orbit spaces that arise from memories whose little groups are connected subgroups of $\grp{SL}(2,\bb{C})$ since these are the cases that will generically arise in scattering. For completeness, an analysis of non-connected subgroups is provided in appendix \ref{sec:non-connected}. Thus, the Hilbert spaces $\Fock_{\textrm{DI},\Delta}$ consist of a large class of asymptotic states with well-defined BMS charges. 

Finally, we compare and contrast the approach presented in this paper to other possible approaches in obtaining representations of the BMS group with memory. We first recall that there are at least two approaches to building the standard (zero memory) Fock space $\Fock_{0}$. One could either seek an irreducible representation of the field algebra which contains the invariant vacuum state or one could equivalently build $\Fock_{0}$ from the irreducible representations of the Poincaré group found by Wigner \cite{Wigner}. Thus, there is a clear compatibility between the irreducible representations of the quantum fields and the irreducible representations of the Poincaré group. This compatibility fails for massless fields. The irreducible representations of the quantum fields are the memory Fock representations $\Fock_{\Delta}$ which do {\em not} have a well-defined action of the BMS group (or any Poincaré subgroup). Furthermore, while irreducible representations of the BMS group were classified by McCarthy \cite{McCarthy1,McCarthy2,McCarthy-nucl}, it is not known to us whether any representation of the field algebra that admits a unitary representation of the (infinite dimensional) BMS group can be decomposed in terms of McCarthy's irreducible representations.\footnote{A theorem by Mautner \cite{Mautner} shows that any representation of a \emph{locally compact} group can be written as a direct integral over irreducible representations. However, the BMS group is not locally compact, so one cannot argue that any representation is decomposable in terms of irreducible ones. A partial relationship between the representations of the asymptotic field algebra and McCarthy's irreducible representations of the BMS group is shown in \cite{KP-BMS-particles}.} Therefore, the approach taken in this paper is to consider representations of the {\em full} algebra of fields and BMS/Poincaré charges as opposed to seeking representations of any subalgebra. 

The structure of the rest of the paper is as follows. In \cref{sec:asymp-quant} we give a brief review of the asymptotic quantization of general relativity at null infinity. In \cref{sec:asympquant} we recall the asymptotic algebra of radiative fields, in \cref{sec:asympquantBMS} we extend this algebra to include the BMS charges, and in \cref{sec:memFock} we recall the construction of Fock spaces with definite memory. In \cref{sec:Lorentzorb} we construct representations of the radiative fields with well-defined BMS charges. In \cref{subsec:gencon}, we provide the general construction of the Hilbert spaces $\Fock_{\textrm{DI},\Delta}$ as well as the explicit action of the Lorentz transformations and charges. In \cref{sec:memory-orbits} we summarize the classification of the possible orbit spaces (obtained in appendix \ref{sec:weighted-func}) and emphasize which orbit spaces are most relevant for scattering theory. 

In appendix \ref{sec:weighted-func} we obtain mathematical results on ``conformal-spin weighted functions'' necessary for the main arguments of this paper. We note that the results presented in this appendix are more general than what is strictly necessary in the paper and may be of mathematical interest in its own right.  In appendix \ref{sec:little-groups-orbit} we classify the connected sub-groups of the Lorentz group that preserve such conformal-spin weighted functions as well as their corresponding orbit spaces. In appendix \ref{sec:orbit-stuff}, we construct the invariant measures on these spaces and the results of these sections are summarized in table \ref{tab:orbits-table}. Finally, in appendix \ref{sec:non-connected} we summarize the results on non-connected groups. 

\section{Asymptotic quantization of general relativity at null infinity}\label{sec:asymp-quant}

In this section, we review the asymptotic quantization of vacuum general relativity at null infinity. In \cref{sec:asympquant}, we will first review the $\ast$-algebra quantization of the asymptotic fields. In \cref{sec:asympquantBMS}, we review the enlargement of this algebra to include the BMS charges and fluxes. In \cref{sec:memFock}, following Ashtekar \cite{,asymp-quant}, we construct a large supply of states of definite memory and illustrate the failure of these states to be extendable to states with well-defined BMS charges. We will primarily focus on the quantization of ``outgoing'' fields at $\scri^{+}$, however an analogous construction of the algebra as well its representations can be done for the ``incoming'' data at $\scri^{-}$. We will therefore not distinguish between past and future null infinity, denoting it as $\scri$, in the following. For further details on the asymptotic quantization of fields at null infinity as well as for massive fields at timelike infinity, we refer the reader to \cite{asymp-quant,Campiglia:2015qka,PSW-IR}. The reader familiar with the material in these references may skip to \cref{sec:Lorentzorb}.

\subsection{Algebra of asymptotic quantum fields}
\label{sec:asympquant}
We now describe the quantization of the radiative degrees of freedom at null infinity. The radiative degrees of freedom of the gravitational field are encoded in the shear tensor $\sigma_{AB}(x)$ on $\scri$ where $x\in \scri$. Here, and in the following, \(u\) denotes the retarded/advanced time coordinate on \(\scri\) and \(x^A\) are some choice of coordinates on \(\bb S^2\) and so $x=(u,x^{A})$. The capital indices \(A,B,\ldots\) are abstract indices of tensor fields on \(\bb S^2\). The Bondi News tensor is defined as 
\begin{equation}
N_{AB}\defn 2\partial_{u}\sigma_{AB}.
\end{equation}
The asymptotic quantization algebra $\mathscr{A}$ is defined as the unital $\ast$-algebra generated by the News smeared with the real-valued test tensor $s^{AB}$ on $\scri$ 
\begin{equation}
\label{eq:N(s)}
\op{N}(s)\defn \int_{\scri}d^{3}x~\op{N}_{AB}(x)s^{AB}(x),
\end{equation}
its formal adjoint $\op{N}(s)^{\ast}$ and the identity $\op{1}$ factored by the following relations:
\begin{condlist}[label=(A.{\Roman*})]
\label{GRalg}
\item $\op{N}(c_{1}s_{1}+c_{2}s_{2})=c_{1}\op{N}(s_{1})+c_{2}\op{N}(s_{2})$ for any $s^{AB}_{1},s^{AB}_{2}$ and any $c_{1},c_{2}\in \mathbb{R}$, \label{A1}
\item $\op{N}(s)^{\ast}=\op{N}(s)$ for all  $s^{AB}$, i.e., the field is Hermitian \label{A2}
\item $[\op{N}(s_{1}), \op{N}(s_{2})]=-64\pi^{2}i\Omega_{\scri}(s_{1},s_{2})\op{1}$, \label{A3}
\end{condlist}
where $\Omega_{\scri}$ is the symplectic product of radiative data at null infinity 
\be
\Omega_{\scri}(s_{1},s_{2})= - \frac{1}{8\pi} \int_{\scri} d^{3}x~\lb[ s_{1AB} \partial_u s_2^{AB} - s_{2AB} \partial_u s_1^{AB}  \rb].
\ee
\ref{A1} expresses the fact that $\op{N}(s)$ is linear in the test function consistent with \cref{eq:N(s)}.  \ref{A2} represents that $\op{N}(s)$ is Hermitian. Finally, \ref{A3} represents the canonical commutation relations for the initial data on $\scri$ which arise from the Poisson bracket relations at null infinity \cite{asymp-quant}. States on $\mathscr{A}$ are positive linear functionals on the algebra, i.e., a state $\omega$ is a map $\omega:\mathscr{A}\to \mathbb{C}$ such that $\s(\op{O}\op{O}^{\ast})\geq 0$ for all $\op{O}\in \mathscr{A}$. Since any element of $\mathscr{A}$ consists of sums of products of the local fields $\op{N}(s)$, the specification of a state is equivalent to a specification of its ``smeared $n$-point correlation functions'' $\s(\op{N}(s_{1})\dots \op{N}(s_{n}))$. This general definition of a state admits many states with singular ultraviolet behavior --- too singular for the radiated energy to null infinity to be finite. Thus, we must impose a further restriction on the states that they ``look like'' the vacuum at short distances (see, e.g., \cite{Kay:1988mu}). While such a condition is necessary, it is not normally explicitly stated since the vacuum state as well as a dense set of states in the standard Fock space satisfy the required ultraviolet conditions. However, as highlighted in the introduction, to accommodate all states that arise in scattering theory, one must go beyond the standard Fock space and so it is essential that we explicitly impose this condition.  This condition is known as the \emph{Hadamard condition} which for the $2$-point function of any state $\omega$ is
\begin{equation}
\label{eq:hadform}
\omega(\op{N}_{AB}(x_{1})\op{N}_{CD}(x_{2}))=-8\frac{(q_{A(C}q_{D)B}-\frac{1}{2}q_{AB}q_{CD})\delta_{\mathbb{S}^{2}}(x_{1}^{A},x_{2}^{B})}{(u_{1}-u_{2}-i0^{+})^{2}} + S_{ABCD}(x_{1},x_{2})
\end{equation}
where all “unsmeared” formulas here and below should be interpreted as holding distributionally. The tensor $S_{ABCD}$ is a state-dependent, bi-tensor on $\scri$ that is symmetric and trace-free in pair of indices $A,B$ and $C,D$ and is symmetric under the simultaneous interchange of $x_{1}$ and $x_{2}$ and the pair of indices $A,B$ with the pair $C,D$. Additionally we require that $S_{ABCD}$ as well as the connected $n$-point functions for $n\neq 2$ of a Hadamard state are sufficiently regular in the sense that they are smooth in retarded/advanced time, decay as  $O((\sum_{i}u_{i}^{2})^{-1/2-\epsilon})$ for some $\epsilon>0$ and are square-integrable in angles. These regularity requirements are sufficient to ensure that all states have finite energy and angular momentum. An important class of states is the \emph{Gaussian states} whose $1$- and $2$-point correlation functions determine all higher-order correlation functions just as in the $n$-th moment of a Gaussian distribution. In other words, such states have vanishing ``connected $n$-point functions'' for $n>2$. Indeed, the vacuum state $\s_{0}$ corresponds to a Gaussian state with vanishing $1$-point function and the $2$-point function is
\begin{equation}
\label{eq:vac2pt}
\omega_{0}(\op{N}_{AB}(x_{1})\op{N}_{CD}(x_{2})) = -8\frac{(q_{A(C}q_{D)B}-\frac{1}{2}q_{AB}q_{CD})\delta_{\mathbb{S}^{2}}(x_{1}^{A},x_{2}^{B})}{(u_{1}-u_{2}-i0^{+})^{2}}
\end{equation}
given by \cref{eq:hadform} with $S_{ABCD}=0$. 

\subsection{Extension of the asymptotic algebra to include the BMS charges}
\label{sec:asympquantBMS}

We now review the extension of $\ms{A}$ to the algebra $\ms{A}_{\Delta,\textrm{Q}}$ which includes the memory and the BMS charges at spatial infinity. We refer the reader to, e.g., \cite{PSW-IR} for further details on this construction. In general relativity, the gauge symmetries correspond to diffeomorphisms. However, the transformations that preserve the asymptotic structure of the metric but do not vanish at infinity can have a non-trivial action on the phase space of general relativity. In four dimensions, the group of such diffeomorphisms (modulo the group of diffeomorphisms with a trivial action on phase space) is known as the BMS group \cite{PhysRev.128.2851}. Given a choice of retarded time coordinate $u$ along the null generators of $\scri$ and coordinates \(x^A\) on \(\bb S^2\) the BMS transformations are given by
\be\label{eq:bms-coord}
    u \mapsto \varpi_\Lambda(x^A) (u + f(x^A)) \eqsp x^A \mapsto \Lambda^A{}_B x^B
\ee
where \(f(x^A)\) is a function on \(\bb S^2\) of conformal weight \(+1\) parametrizing the BMS supertranslations and \(\Lambda\) parametrizes Lorentz transformations acting on \(\bb S^2\) as conformal transformations. The function \(\varpi_\Lambda(x^A)\) is a strictly positive smooth function on \(\bb S^2\), which is the conformal change in the degenerate metric \(q_{AB} \mapsto \varpi_\Lambda^2 q_{AB}\) under the Lorentz transformation. The explicit form of the Lorentz transformation and \(\varpi_\Lambda\) in stereographic coordinates is given in \cref{eq:z-transform,eq:omega-h-defn}. Infinitesimal BMS transformations are generated by the vector field
\begin{equation}
\xi^{\mu} = \lb( f + \frac{1}{2}u\mathscr{D}_{A}X^{A} \rb) \bigg(\pd{u}\bigg)^{\mu} + X^{A}\bigg( \pd{x^A} \bigg)^{\mu}
\end{equation}
The vector $X^{A}$ is tangent to the constant $u$ cross-sections of $\scri$ and is a conformal Killing vector field on $\bb{S}^{2}$.  Thus, given a choice of retarded time, a particular BMS transformation corresponds to a specification of the pair $(f,X)$. The transformations corresponding to $(f,0)$ are the \emph{supertranslations} and if $f$ is a linear combination of only $\ell=0,1$ spherical harmonics then it corresponds to an ordinary translation. The transformation generated by $(0,X)$ are Lorentz transformations.\footnote{The decomposition of a BMS transformation into supertranslation and a Lorentz transformation depends on the choice of cross-section of constant retarded time coordinate $u$. This is due to the fact that the BMS group is the semidirect product of the Lorentz group with the abelian group of supertranslations, and thus there is no unique choice of Lorentz subgroup within the BMS group.}

The action of an infinitesimal, full BMS transformation on the shear is given by
\begin{equation}
\label{eq:sigmaBMS}
\sigma_{AB} \mapsto \sigma_{AB} + \epsilon \bigg[\tfrac{1}{2}(f+\tfrac{1}{2}u\ms{D}_{C}X^{C})N_{AB} + \lb(\ms{D}_{A}\ms{D}_{B}-\tfrac{1}{2}q_{AB}\ms{D}^{2}\rb) f + \Lie_{X}\sigma_{AB}-\tfrac{1}{2}(\ms{D}_{C}X^{C})\sigma_{AB}\bigg]
\end{equation}
and the corresponding action on $N_{AB}$ can be obtained using the fact that $N_{AB}=2\partial_{u}\sigma_{AB}$. The BMS charges correspond to the observables which generate the above BMS transformation. The BMS charge which generates Lorentz transformations $(0,X)$ in \cref{eq:sigmaBMS} is given by \cite{GPS}
\begin{equation}
{\mc Q}_{i^0} (X) = \frac{1}{16\pi} \int_\scri du d\Omega~ N^{AB} \bigg[ \frac{1}{4} ( u \ms D_C X^C) N_{AB}+ \Lie_X \sigma_{AB}- \tfrac{1}{2} (\ms D_C X^C) \sigma_{AB}\bigg].
\end{equation}
The supertranslation charge ${\mc Q}^{\GR}_{i^0} (f) $ generates the transformation $(f,0)$. It will be useful to separate off the ``time-independent'' piece of the supertranslation in \cref{eq:sigmaBMS} 
\begin{equation}
\label{eq:sigmasuper}
    \sigma_{AB} \mapsto \sigma_{AB} + \epsilon \lb( \ms{D}_{A}\ms{D}_{B}-\tfrac{1}{2}q_{AB}\ms{D}^{2} \rb)f.
\end{equation}
The observable on $\scri$ which generates this transformation is the gravitational memory observable\footnote{The memory can be decomposed as $\Delta_{AB}=\big(\ms{D}_{A}\ms{D}_{B}-(1/2)q_{AB}\ms{D}^{2}\big)\alpha + \varepsilon_{(A}{}^{C}\ms{D}_{B)}\beta$ where $\alpha$ and $\beta$ are functions on $\bb{S}^{2}$ corresponding to the ``electric'' and ``magnetic'' parity parts respectively. \Cref{eq:memel} concerns only the electric parity part of the memory. All of the constructions of this paper can be straightforwardly extended to the magnetic parity part. While the magnetic parity memory can occur for certain physical processes (see \cite{Satishchandran_2019} for an explicit example) it does not arise for scattering with incoming data with vanishing memory. } 
\begin{equation}
\label{eq:memel}
    \Delta(f)\defn \frac{1}{2}\int_\scri dud\Omega~N_{AB}(u,x^{C})\ms{D}^{A}\ms{D}^{B}f(x^{C})
\end{equation}
which, we note, vanishes if $f$ is a linear combination of $\ell=0,1$ spherical harmonics. The observable on $\scri$ which generates the ``time-dependent'' part of the supertranslation $(f,0)$ in  \cref{eq:sigmaBMS} is
\begin{equation}
\label{eq:nullmem}
\mc{J}(f) = \frac{1}{32\pi}\int_{\scri}dud\Omega~fN_{AB}N^{AB}
\end{equation}
and the full supertranslation charge is given by 
\begin{equation}
\mc{Q}_{i^{0}}(f) = \mc{J}(f) + \frac{1}{8\pi}\Delta(f). 
\end{equation}
Finally, the full BMS charge is $\mc{Q}_{i^{0}}(f,X)=\mc{Q}_{i^{0}}(f)+\mc{Q}_{i^{0}}(X)$.

The algebra $\ms{A}_{\Delta,\textrm{Q}}$ is generated by including in $\ms{A}$ the additional elements $\op{\mc{Q}}_{i^{0}}(f),\op{\mc{Q}}_{i^{0}}(X)$ and $\op{\Delta}(f)$ factored by the following commutation relations. The commutation relations of $\op{\mc{Q}}_{i^{0}}(f,X) = \op{\mc{Q}}_{i^{0}}(f)+\op{\mc{Q}}_{i^{0}}(X)$ with the radiative fields are given by 
\be
    \lb[ \op{\mc Q}_{i^0}(f,X), \op N(s) \rb] = i \op N(\delta_{(f,X)}s)
\ee
where
\be
    \delta_{(f,X)} s^{AB} & = \lb( f + \tfrac{1}{2} u \ms D_C X^C \rb) \partial_u s^{AB} + \Lie_X s^{AB} + \tfrac{1}{2} (\ms{D}_{C} X^{C}) s^{AB}
\ee
We also have that 
\begin{align}
    &[\op{\mc{Q}}_{i^{0}}(f_{1}),\op{\mc{Q}}_{i^{0}}(f_{2})]=0,  \quad [\op{\mc{Q}}_{i^{0}}(X_{1}),\op{\mc{Q}}_{i^{0}}(X_{2})] = i \op{\mc{Q}}_{i^{0}}([X_{1},X_{2}]) \nonumber \\
    &\quad \quad [\op{\mc{Q}}_{i^{0}}(X),\op{\mc{Q}}_{i^{0}}(f)] = i \op{\mc{Q}}_{i^{0}}\lb( \Lie_X f-\tfrac{1}{2}(\ms{D}_{A}X^{A})f \rb)
\end{align}
where $[X_{1},X_{2}]$ is the Lie bracket of $X_{1}^{A}$ and $X_{2}^{A}$. The memory observable has vanishing Poisson brackets with the News, itself and the supertranslation charge 
\begin{equation}
    [\op{\Delta}(f),\op{N}(s)] = [\op{\Delta}(f_{1}),\op{\Delta}(f_{2})] = [\op{\mc{Q}}_{i^{0}}(f),\op{\Delta}(f)] = 0 
\end{equation}
however, the memory is not Lorentz invariant 
\begin{equation}
\label{eq:Lorentzmem}
    [\op{\mc{Q}}_{i^{0}}(X),\op{\Delta}(f)] = i \op{\Delta}\lb(\Lie_X f-\tfrac{1}{2}(\ms{D}_{C}X^{C})f\rb)
\end{equation}
and so does not commute with the Lorentz charge. The algebra generated by $\op{N}$, $\op{\mc{Q}}_{i^{0}}$ and $\op{\Delta}$ yields the desired extended algebra $\ms{A}_{\textrm{$\Delta$,Q}}$. 

\subsection{Memory Fock representations of the asymptotic field algebra}
\label{sec:memFock}

In the previous subsections, we reviewed the construction of the asymptotic local field algebra $\ms{A}$ as well as its extension $\ms{A}_{\textrm{$\Delta$,Q}}$ to include the memory and the generators of the BMS charges. The physical states of quantum gravity correspond to states with well-defined energy, supermomentum and angular momentum and are therefore states on the extended algebra. The explicit construction of states on $\ms{A}_{\textrm{$\Delta$,Q}}$ is the primary aim of this paper. To achieve this we will first review the construction of a large supply of states with ``definite memory'' on $\ms{A}$ \cite{PhysRevLett.46.573,asymp-quant}. However, as we will see, these states will generally not have a well-defined action of the Lorentz group and therefore will not be extendable to states on $\ms{A}_{\textrm{$\Delta$,Q}}$. In \cref{sec:Lorentzorb} we will, appropriately re-assemble these states into physical states on the extended algebra. 

The vacuum state $\omega_{0}$, however, is an important exception to the above caveats since it is annihilated by {\em all} of the elements of the extended algebra, i.e., $\s_{0}(\op O \op{\Delta})=\s_{0}(\op O \op{\mc{Q}}_{i^{0}})=\s_{0}(\op{\Delta} \op O)=\s_{0}(\op{\mc{Q}}_{i^{0}} \op O)=0$ for any $\op O \in \ms{A}_{\textrm{out,Q}}$. The corresponding Hilbert space built from ``acting'' the News operators $\op{N}(s)$ on the vacuum\footnote{More precisely, given an algebraic state $\s$ one can use this state to define an inner-product on the algebra whose completion yields a Hilbert space. This is known as the \emph{GNS construction} \cite{GNS1,GNS2} (see, e.g., \S 3 of \cite{PSW-IR}).} is the Fock space $\Fock_{0}$
\be 
    \Fock_{0}= \mathbb{C}\oplus \bigoplus_{n\geq 1}\underbrace{\big(\Hilb_{0}\otimes_{S}\dots \otimes_{S}\Hilb_{0}\big)}_{n\text{ times}}. 
\ee 
where $\otimes_{S}$ is the symmetrized tensor product and the one-particle Hilbert space $\Hilb_{0}$ is the completion of the space of positive frequency test function $s^{AB}$ with respect to the inner product
\begin{equation}
\label{eq:innerprod}
\braket{s_{1}|s_{2}}_{0}\defn \s_{0}(\op{N}(s_{1})\op{N}(s_{2}))= 16\pi \int_{0}^{\infty}\omega d\omega \int_{\bb{S}^{2}}d\Omega ~\bar{s_{1}^{AB}(\omega, x^A)}s_{2,AB}(\omega, x^A)
\end{equation}
and  $s^{AB}(\omega, x^A)$ is the Fourier transform of the test tensor in \(u\) with frequency \(\omega\). 

As is well known, the standard Fock space of gravitons yields a dense set of Hadamard states with finite energy, (super)momentum, and angular momentum.\footnote{The Lorentz charges can be represented on the zero memory Fock space \(\Fock_0\) and can be used to define the helicity of the graviton states. The helicity of one graviton states in \(\Fock_0\) can be computed to be \(\pm 2\) as expected; see \cite{asymp-quant} for details.} However, the standard Fock space does not include all of the states physically relevant for scattering theory. Indeed, since the memory operator $\op{\Delta}$ annihilates $\ket{\s_{0}}$ and the fact that $\op{\Delta}$ commutes with $\op N(s)$ implies that {\em all} states in $\Fock_{0}$ have zero memory. Since the memory is not generically conserved, an in-state in $\Fock_{0}$ --- with well-defined energy and angular momentum --- will scatter to an out-state with nonzero memory. 

Nevertheless, a large supply of states with non-vanishing memory can be straightforwardly constructed following the procedure in \cite{asymp-quant}. Let $\nu_{AB}(x)$ be a classical News tensor at null infinity which is regular as explained below \cref{eq:hadform} and has classical memory 
\begin{equation}
    \Delta(\nu,f) = \int_{\scri} dud\Omega~\nu_{AB}(u,x^{C}) \ms D^A \ms D^B f(x^C).
\end{equation}
Consider the map \(\mf a_\nu\) on operators defined a follows
\begin{equation}
    \mf{a}_{\nu}[\op{N}(s)] = \op{N}(s) + \nu(s)\op{1} \eqsp  \mf{a}_{\nu}[\op{\Delta}(f)] = \op{\Delta}(f) + \Delta(\nu,f)\op{1}.
\end{equation}
Then, extending this map so that it preserves the identity \(\1\), products of operators and the \(*\)-relation we get an automorphism of the subalgebra $\ms{A}_{\Delta}$ generated by \(\1\), $\op{N}(s)$ and $\op{\Delta}(f)$. Using this automorphism, we can define a new state $\s_{\nu}$ whose correlation functions are given by 
\begin{equation}
    \s_{\nu}(\op O) = \s_{0}\lb(\mf{a}_{\nu}[\op O]\rb)
\end{equation}
for any $\op O \in \ms{A}_{\Delta}$. The state $\s_{\nu}$ is a Gaussian state where its $n$-point functions are shifted by the classical solution $\nu_{AB}(x)$, in particular, it has a definite memory given by \(\Delta(\nu,f)\). The GNS construction over this ``shifted vacuum'' i.e. acting with $\op N(s)$ yields a Fock representation $\Fock_{\nu}$. If $\Delta(\nu,f) = 0$ then the automorphism \(\mf a_\nu\) can be represented in the standard Fock space \(\Fock_0\) by a unitary operator and the new state \(\s_\nu\) is a coherent state in \(\Fock_0\). Other the other hand, if \(\Delta(\nu,f)\) is chosen to be non-vanishing, then \(\mf a_\nu\) cannot be unitarily implemented in \(\Fock_0\) and the new Fock space \(\Fock_\nu\) is \emph{unitarily inequivalent} to the standard Fock space $\Fock_{0}$. The choice of ``vacuum'' state $\s_{\nu}$ of $\Fock_{\nu}$ is not unique and one could have chosen a different News tensor $\nu'_{AB}$ with the same memory, i.e., $\Delta(\nu', f) = \Delta(\nu, f)$ for all \(f\). Nevertheless, as long as $\nu'_{AB}$ is sufficiently regular\footnote{Two Fock representations $\Fock_{\nu}$ and $\Fock_{\tilde{\nu}}$ with radiation fields $\nu_{AB}=2\partial_{u}\sigma_{AB}$ and $\tilde{\nu}_{AB}=2\partial_{u}\tilde{\sigma}_{AB}$ are unitarily equivalent if and only if the norm of $\sigma_{AB}-\tilde{\sigma}_{AB}$ (defined by \cref{eq:innerprod}) is finite. Even if the radiation fields have the same memory, this norm may diverge if they are not sufficiently regular.}, then the two representations will be unitarily equivalent. Thus the unitary equivalence class of all $\Fock_{\nu}$ with sufficiently regular representative $\nu_{AB}$ with memory $\Delta_{AB}$ can be labeled by the memory of the representative --- i.e. we may label the ``memory representations'' as $\Fock_{\Delta}$ rather than $\Fock_{\nu}$. Each memory Fock representation $\Fock_{\Delta}$ is unitarily inequivalent to $\Fock_{\Delta'}$ for $\Delta\neq \Delta'$. 

While the above construction yields a large class of states with memory, these states cannot be extended to physical states on the full algebra $\ms{A}_{\Delta,\textrm{Q}}$ with well-defined BMS charges. Let's first consider the action of supertranslations. The action of supertranslations does not change the memory of the radiation field and so the supertranslation of any chosen classical News $\nu_{AB}$ yields another News $\nu'_{AB}$ with the same memory. However, as we have already noted, the representation formed from $\nu_{AB}$ and $\nu'_{AB}$ with the same memory are unitarily equivalent. For sufficiently regular $\nu_{AB}$, this unitary map is strongly continuous, and therefore the Fock space $\Fock_{\Delta}$ has a well-defined action of the supertranslation charge $\op{\mc{Q}}_{i^{0}}(f)$. Indeed, from \cref{eq:nullmem}, finiteness of the supertranslation charges places a necessary condition that the allowed memories must be square-integrable on $\bb{S}^{2}$ as noted in \cref{eq:memL2} of the introduction. We emphasize that while the supertranslation charges $\op{\mc{Q}}_{i^{0}}(f)$ have a well-defined action in any memory Fock space, its spectrum is only bounded below by zero in any $\Fock_{\Delta}$ and only has a vanishing eigenvalue in the standard Fock space $\Fock_{0}$ where the eigenvector is the vacuum $\ket{\s_{0}}$ (see also \cite{asymp-quant}). 

However, the memory Fock representations do not have a well-defined unitary action the Lorentz group \cite{asymp-quant}. Since the memory is not Lorentz-invariant, unless it vanishes, the action of a general Lorentz transformation on \(\Fock_\Delta\) will give a different memory and map to a unitarily inequivalent memory Fock space. Thus, the infinitesimal generator of Lorentz transformations, i.e. the Lorentz charges \(\op{\mc Q}_{i^0}(X)\), are not defined on any \(\Fock_\Delta\) when \(\Delta \neq 0\). Another way to see this is to note that, all states in the Fock representations $\Fock_{\Delta}$ are eigenstates of the memory operator $\op{\Delta}$, and thus any operator defined on $\Fock_{\Delta}$ commutes with \(\op\Delta\). By the commutator \cref{eq:Lorentzmem}, it immediately follows that, the Lorentz charges do not commute with memory, in general, and therefore the Lorentz charges $\op{\mc{Q}}_{i^{0}}(X)$ cannot be represented on the memory Fock space $\Fock_{\Delta}$ as densely defined operators.

However, for any given memory $\Delta_{AB}$, consider the {\em maximal} subgroup of the full Lorentz group that leaves the memory $\Delta_{AB}(x^{A})$ invariant. We call this the \emph{little group} $\grp{L}_{\Delta}$ of the chosen memory. We note that a general memory tensor need not have any non-trivial little group. If $\nu_{AB}(x)$ is a News tensor with memory $\Delta_{AB}$ then the Lorentz transformed News tensor \(\nu'_{AB} = (\Lambda \nu)_{AB}\) with \(\Lambda \in \grp L_\Delta\) will have the same memory. By arguments similar to those given above for supertranslations, we can conclude that the little group \(\grp L_\Delta\) can be represented by strongly continuous unitaries on \(\Fock_\Delta\). Thus, the charge operators \(\op{\mc Q}_{i^0}(X)\) can be represented as densely defined operators on \(\Fock_\Delta\) if and only if \(X\) is a generator of the little group \(\grp L_\Delta\), but the charge operators generating Lorentz transformation that are not in the little group cannot be represented. Consequently, even though the memory Fock spaces \(\Fock_\Delta\) provide us a large supply of states with memory, they cannot be extended to the full algebra \(\ms A_{\Delta,\rm Q}\). 

\section{Representations of the BMS group with memory}
\label{sec:Lorentzorb}

In the previous section, we reviewed how to construct a large class of states with memory. However, all of the states constructed could not be extended as physical states the action of the Lorentz charges cannot be defined. In \cref{subsec:gencon} we provide the general construction of a large supply of states which have a well-defined action of all of the BMS charges by constructing direct integral representations over the Lorentz orbit space of a chosen memory. In particular, we give the explicit action of a finite Lorentz transformation as well as the action of all Lorentz charges in these representations. Finally, in \cref{sec:memory-orbits}, we summarize the results presented in appendix \ref{sec:orbit-stuff} on the classification of the orbit spaces that arise in scattering theory. 

\subsection{General construction of representations of the BMS group with memory}
\label{subsec:gencon}

Recall that, for any Fock representation $\Fock_{\Delta}$ with non-vanishing memory, one can straightforwardly impose suitable restrictions on the regularity of these states such that they have a well-defined action of all supertranslation charges --- the memories must be square-integrable on the $2$-sphere. However, one cannot extend these states to have a well-defined action of the Lorentz group. The most one could do was have a well-defined action of a subgroup of the Lorentz group which preserves the memory of the state. 

In this section, we construct separable Hilbert spaces with a continuous unitary action of the full BMS group, including all Lorentz transformations, and with non-zero memory. All the BMS charges will then be well-defined operators on a dense subspace of states. This provides us with a large supply of states with memory and well-defined BMS charges which are suitable for constructing an infrared finite scattering theory that can incorporate the conservation of all BMS charges \cite{Strominger:2013jfa}.

We start with a chosen memory tensor \(\Delta_{AB}\) and the corresponding memory Fock space \(\Fock_\Delta\). As discussed in the previous section, the BMS supertranslations and the subgroup of the Lorentz transformations which preserve the memory, i.e., the little group \(\grp L_\Delta\) are unitarily implemented on \(\Fock_\Delta\). The Lorentz transformations which are \emph{not} unitarily implemented can be parametrized by the quotient\footnote{We emphasize that, in general, \(\orbit_\Delta\) is not a group; in our case, it does have the structure of a smooth finite-dimensional manifold, possibly with boundaries.}
\begin{equation}
\label{eq:ODelta}
    \orbit_{\Delta} \cong \grp{SL}(2,\bb C)/\grp{L}_{\Delta} \,.
\end{equation}
We will sometimes denote the ``orbit space'' as $\ms{O}$ when referring to it abstractly. This quotient consists of equivalence classes of Lorentz transformations modulo the right multiplication by any element of the little group, i.e., \(\Lambda' \sim \Lambda\) if and only if there exists a \(\Lambda'' \in \grp L_\Delta\) such that \(\Lambda' = \Lambda \Lambda''\). Thus, the entire little group \(\grp L_\Delta\) lies in the equivalence class of the identity transformation, and the rest of the elements of \(\orbit_\Delta\) parametrize the Lorentz transformations not in the little group. If we act with this equivalence relation on the memory we get \((\Lambda'\Delta)_{AB} = (\Lambda \Delta)_{AB}\) since \(\Lambda''\) preserves the memory by definition of the little group. Thus the quotient space \(\orbit_\Delta\) also parametrizes all the possible memories we can obtain by acting on the initially chosen one, \(\Delta_{AB}\), by all Lorentz transformations. \(\orbit_\Delta\) is the \emph{orbit space} of \(\Delta_{AB}\) under Lorentz transformations. We denote a point in the orbit space by \(y \in \orbit_\Delta\), and let \(y=y_0\) denote the point corresponding to the equivalence class of the identity transformation. Then, with each point \(y \in \orbit_\Delta\) we can associate a memory \(\Delta(y)\) which is obtained from the initially chosen memory \(\Delta\) by acting with a Lorentz transformation in the equivalence class corresponding to \(y\) and, the initially chosen memory is \(\Delta = \Delta(y_0)\). Similarly, with each point \(y \in \orbit_\Delta\) we can associate the entire Fock space \(\Fock_{\Delta(y)}\), i.e., we have a collection of memory Fock spaces parametrized by a point in the orbit space \(\orbit_\Delta\). Further, if \(\Delta(y) = \Lambda \Delta\) where \(\Lambda\) is in the equivalence class at \(y\), then the little group of \(\Delta(y)\) is given by the conjugation \(\grp{L}_{\Delta(y)} = \Lambda \grp{L}_\Delta \Lambda^{-1}\).

Now we ``stitch'' together these memory Fock spaces \(\Fock_{\Delta(y)}\) to get a representation of the full Lorentz group. In order to obtain a well-defined action of the {\em finite} Lorentz transformations one could consider the (uncountable) direct sum of over all the memory Fock spaces \(\Fock_\Delta(y)\), i.e.,
\begin{equation}
\label{eq:DS}
\Fock_{\textrm{DS},\Delta}=\bigoplus_{y \in \orbit}\Fock_{\Delta(y)}
\end{equation}
 where the subscript ``\textrm{DS}'' refers to the direct sum. However, there are at least two serious issues with these representations. The first is that, since \cref{eq:DS} is a direct sum over all points in the orbit space $\orbit$ this Hilbert space is clearly nonseparable.\footnote{The direct sum would be a separable Hilbert space if the orbit space is a countable set, but such orbits will not arise in our case (see  \cref{sec:memory-orbits} and appendix \ref{sec:little-groups-orbit}).} Further, since even a ```small'' Lorentz transformation, will map a vector in $\Fock_{\Delta}$ into an orthogonal sector in the direct sum, the Lorentz transformations cannot be strongly continuous. Therefore the Lorentz charges (e.g., the angular momentum) which are the infinitesimal generators of Lorentz transformations, cannot be defined.\footnote{This situation is analogous to what occurs when one considers the Hilbert space in one-dimensional Schrödinger quantum mechanics to be the (non-separable) direct sum $\bigoplus_{x}\ms{H}_{x}$ where $\ms{H}_{x}$ is the Hilbert space of eigenstates of the position operator with eigenvalue $x$. This Hilbert space has a well-defined action of translations but it is not strongly continuous and therefore the momentum operator cannot be defined.}

 To obtain representations under which the BMS group has a strongly continuous, unitary action one must ``integrate'' instead of ``sum'' the memory Fock spaces $\Fock_{\Delta(y)}$ over $\ms{O}_{\Delta}$. As summarized in \cref{sec:memory-orbits} and shown in appendix \ref{sec:orbit-stuff}, each orbit space $\ms{O}_{\Delta}$ of interest has a unique, Lorentz invariant measure $\mu_{\ms{O}}(y)$. One can then construct the direct integral Hilbert space $\Fock_{\textrm{DI},\Delta}$ from the memory Fock spaces $\Fock_{\Delta(y)}$ in the following manner: An element $\ket{\Psi}\in \Fock_{\textrm{DI},\Delta}$ is given by specifying a measurable family of vectors $\ket{\psi(y)} \in  \Fock_{\Delta(y)}$ for all $y\in \ms{O}_{\Delta}$. Two states $\ket{\Psi}$ and $\ket{\Psi'}$ are considered equivalent if their family of vectors $\ket{\psi(y)}$ and $\ket{\psi'(y)}$ differ only on a set of measure zero with respect to $\mu_{\ms{O}}(y)$. The states in \(\Fock_{\rm{DI},\Delta}\) are required to have finite norm
\begin{equation}
    \norm{\Psi}^2 = \int_{\ms{O}}d\mu_{\ms{O}}(y)~\braket{\psi(y) | \psi(y)}_{\Delta(y)} < \infty
 \end{equation}
 and the inner product of two states is then given by 
\begin{equation}
    \braket{\Psi|\Psi^{\prime}}=\int_{\ms{O}}d\mu_{\ms{O}}(y)~\braket{\psi(y) | \psi'(y)}_{\Delta(y)}.
 \end{equation}
The inner product on the right-hand side of the above two equations is the inner product evaluated on \(\Fock_{\Delta(y)}\) at the point \(y \in \orbit_\Delta\). We denote the resulting direct integral Hilbert space as
\begin{equation}
    \Fock_{\textrm{DI},\Delta}=\int_{\ms{O}}^\oplus d\mu_{\ms{O}}(y)~\Fock_{\Delta(y)}
\end{equation}
where the subscript ``\textrm{DI},$\Delta$'' refers to the fact that this is a direct integral representation associated to the choice of memory $\Delta_{AB}$. This yields a separable Hilbert space and, as we will show below, admits a strongly continuous unitary representation of the full BMS group, i.e. all BMS charges have a well-defined action on $\Fock_{\textrm{DI},\Delta}$. We note that, two representations $\Fock_{\textrm{DI},\Delta}$ and $\Fock_{\textrm{DI},\Delta'}$ are unitarily equivalent if there exists a Lorentz transformation $\Lambda \in \grp{SL}(2,\bb C)$ such that $\Lambda \Delta = \Delta'$. Since $\Delta_{AB}$ is an element of the (infinite-dimensional) space of square-integrable, symmetric, trace-free tensors on $\bb{S}^{2}$ and the orbit spaces are finite-dimensional it's clear that the collection of inequivalent Hilbert spaces $\Fock_{\textrm{DI},\Delta}$ is uncountably infinite. Therefore, this construction yields a large supply of physical Hadamard states on $\ms{A}_{\Delta,\textrm{Q}}$.

We now obtain the explicit action of the extended algebra \(\ms A_{\Delta,\rm Q}\) including the BMS charges on $\Fock_{\textrm{DI},\Delta}$. We first note that any operator $\op{O}$ that has a well-defined action on each of the memory Fock spaces $\Fock_{\Delta(y)}$ also has a well-defined action on $\Fock_{\textrm{DI},\Delta}$. Thus, the operators $\op{N}(s)$, $\op{\Delta}(f)$ and $\op{\mc{Q}}_{i^{0}}(f)$ have a well-defined action on $\Fock_{\textrm{DI},\Delta}$. Therefore, $\Fock_{\textrm{DI},\Delta}$ carries a unitary representation of the supertranslations. Since the measure $\mu(y)$ is Lorentz invariant, the direct-integral Hilbert space also carries a strongly continuous unitary action of the Lorentz group. However, the memory does not commute with the Lorentz charges and so its action non-trivially ``mixes'' the different memory Fock spaces in $\Fock_{\textrm{DI},\Delta}$ at different points of the orbit. We now obtain the explicit unitary representation of the Lorentz group.

To specify the action of the Lorentz group on \(\Fock_{{\rm DI},\Delta}\) it will be useful to pick a family of News tensors \(\nu_{AB}(x;y)\) and a corresponding choice of vacua \(\ket{\s_{\nu(y)}}\) in every Fock space \(\Fock_{\Delta(y)}\). While the family of News tensors is freely specifiable, it will be convenient to choose $\nu_{AB}(x;y)$ to depend smoothly on $y \in \orbit_\Delta$. This choice will allow us to straightforwardly define the unitary action of the Lorentz group. Any other choice of a family of News tensors on $\orbit_\Delta$ will be related to this choice by a unitary map, and thus, we may consider such a family without loss of generality.  

We choose this family of News tensors as follows. The Lorentz group acts transitively on any orbit space, i.e., for any two points $y,y' \in \orbit_\Delta$ there exists a (not necessarily unique) Lorentz transformation which maps $y$ to $y'$. In appendix \ref{sec:orbit-stuff}, we show that there always exists a smooth map $\lambda : \ms{O} \to \grp{SL}(2,\bb{C})$ which maps the point $y_{0}$ to the point $y$,
\begin{equation}
    \lambda(y)y_{0} = y
\end{equation}
Recall that the point \(y_0\) corresponds to the initial choice of memory. Thus, starting with a News tensor $\nu_{AB}(x;y_{0})$ at $y_{0}$, we can obtain a family of News tensors $\nu_{AB}(x;y)$ for any $y \in \orbit_\Delta$ which smoothly depends on $y$ by transforming \(\nu_{AB}(x;y_0)\) with the Lorentz transformation $\lambda(y)$, i.e.
\begin{equation}
\label{eq:orbit-news-choice}
    \nu_{AB}(x;y) \defn  (\lambda(y) \nu)_{AB}(x;y_{0})
\end{equation}
where on the right-hand side \((\lambda(y) \nu)_{AB}\) denotes the action of a Lorentz transformation on the tensor field \(\nu_{AB}\) obtained by the pullback along the action of the Lorentz transformation on \(\scri\).

We then consider a family of vectors $\ket{\psi(y)}\in \Fock_{\Delta(y)}$ obtained by acting with the News operator $\op{N}$ on the choice of the vacua $\ket{\s_{\nu(y)}}$ 
\be\label{eq:nice-state}
    \ket{\psi(y)} = \sum_{n=0}^\infty~ \int_{\scri^n} d^3x_1 \cdots d^3x_n~ \psi^{A_{1}B_{1}\dots A_{n}B_{n}}(x_1,\ldots,x_n;y) \op N_{A_{1}B_{1}}(x_1) \cdots \op N_{A_{n}B_{n}}(x_n) \ket{\omega_{\nu(y)}}
\ee
where each $\psi^{A_{1}B_{1}\dots A_{n}B_{n}}(x_1,\ldots,x_n;y)$ is a smooth test tensor of compact support (or has sufficiently fast fall-off) on \(\scri^n \times \orbit_\Delta\). The linear span of the states \(\ket{\Psi}\) defined by the family of vectors of the form \cref{eq:nice-state} is dense in $\Fock_{{\rm DI},\Delta}$.

Now, let \(\Lambda \in \grp{SL}(2,\bb C)\) be a given Lorentz transformation. We wish to define its action on states \(\ket{\Psi}\) specified by families of vectors of the form \cref{eq:nice-state}. First, we consider the action of \(\Lambda\) on the chosen family of News tensors \(\nu_{AB}(x;y)\); let \(\nu'_{AB}(x;y)\) be the family of News tensors obtained by acting on the original family \(\nu_{AB}(x;y)\) by \(\Lambda\). Then, the original News tensor at the point \(\Lambda^{-1}y\) gets mapped to the transformed News tensor at the point \(y\) for all \(y \in \orbit_\Delta\). That is we have
\be
    \nu'_{AB}(x;y) = (\Lambda \nu)_{AB}(x; \Lambda^{-1}y).
\ee
Using the choice of News tensors in \cref{eq:orbit-news-choice} we rewrite the right-hand side of the above equation back at the point \(y\) as follows
\be
    \nu'_{AB}(x;y) &= \lb(\Lambda \lambda(\Lambda^{-1}y) \nu\rb)_{AB}(x; y_0) = \lb(\Lambda \lambda(\Lambda^{-1}y) \lambda(y)^{-1} \nu\rb)_{AB}(x; y) \\
    &= \lb(\Lambda'(y)\nu\rb)_{AB}(x; y)
\ee
where we have defined
\be\label{eq:lorentz-little-projection}
    \Lambda'(y) = \Lambda \lambda(\Lambda^{-1}y) \lambda(y)^{-1}
\ee
Note that
\be
    \Lambda'(y)y = \Lambda \lambda(\Lambda^{-1} y) \lambda(y)^{-1} y = \Lambda \lambda(\Lambda^{-1} y) y_0 = \Lambda \Lambda^{-1} y = y
\ee
and thus \(\Lambda'(y)\) preserves the point \(y\), and is an element of the little group \(\grp{L}_{\Delta(y)}\) for every \(y \in \orbit_\Delta\). Thus, the new family of News tensors \(\nu'_{AB}(x;y)\) is related to the original one \(\nu_{AB}(x;y)\) by a Lorentz transformation which lies in the little group of the memory at each point of \(\orbit_\Delta\). It follows that the action of any \(\Lambda \in \grp{SL}(2,\bb C)\) on the corresponding choice of vacua \(\ket{\s_{\nu(y)}}\) is given by a unitary map \(\op U(\Lambda'(y))\) at each \(y \in \orbit_\Delta\),
\be
    \ket{\s_{\nu'(y)}} = \op U(\Lambda'(y)) \ket{\s_{\nu(y)}}
\ee
where \(\Lambda'(y)\) is determined using \cref{eq:lorentz-little-projection}. Recall that each Fock space \(\Fock_{\Delta(y)}\) has a unitary action of the little group \(\grp L_{\Delta(y)}\) and so the above unitary is well-defined. Further, since the \(\lambda(y)\) is smooth in \(y\), the above transformation also preserves the smoothness of the family of News tensors along the orbit.

As the next step, we define how the test tensors in the expression \cref{eq:nice-state} transform under the action of \(\Lambda\). Given the family of test tensors of the original state, we define the transformation so that at the point \(y\) the transformed test tensor is the Lorentz transformation by \(\Lambda\) of the original test tensor at \(\Lambda^{-1}y\), i.e.
\begin{equation}
    {\psi'}^{A_{1}B_{1}\ldots A_{n}B_{n}}(x_{1},\ldots,x_{n} ; y) \defn (\Lambda \psi)^{A_{1}B_{1}\ldots A_{n}B_{n}}(x_{1},\ldots,x_{n};\Lambda^{-1}y).
\end{equation}
Putting together the transformations of both the families of test tensors and choice of vacua we have that under a Lorentz transformation by \(\Lambda \in \grp{SL}(2,\bb C)\), the state \(\ket{\Psi}\) specified by the family of vectors in \cref{eq:nice-state} transforms to \(\ket{\Psi'}\) specified by the family of vectors
\be
\label{eq:LorDI}
    \ket{\psi'(y)} &= \op{U}(\Lambda)\ket{\psi(y)} \\
 &= \sum_{n=0}^{\infty} \int_{\scri^n} d^{3}x_{1}\cdots d^{3}x_{n}~ (\Lambda \psi)^{A_{1}B_{1}\ldots A_{n}B_{n}}(x_{1},\ldots,x_{n};\Lambda^{-1}y)\\  
&\qquad\qquad\qquad \times \op{N}_{A_{1}B_{1}}(x_{1}) \cdots \op{N}_{A_{n}B_{n}}(x_{n}) \op{U}(\Lambda'(y)) \ket{\s_{\nu(y)}}
\ee
where we recall that $\Lambda'(y)\in \grp{L}_{\Delta(y)}$. Using the Lorentz invariance of the measure $\mu(y)$ on $\orbit_\Delta$, it is straightforward to verify that $\op{U}(\Lambda)$ preserves the norm of states of the form \cref{eq:nice-state}. Thus $\op{U}(\Lambda)$ is a unitary map on a dense subspace of $\Fock_{\textrm{DI},\Delta}$, and can be extended to all $\Fock_{\textrm{DI},\Delta}$ by continuity. Therefore, this direct integral over the memory Fock representations is indeed a unitary representation of the full BMS group. 

To show that this representation is indeed strongly continuous we need to obtain the action of the Lorentz charge operator on $\Fock_{\textrm{DI},\Delta}$ on a dense subspace of the Hilbert space. To see this, let $\Lambda(\epsilon)=e^{i\epsilon X}$ be a one-parameter family of Lorentz transformations generated by $X\in \mathfrak{sl}(2,\bb{C})$. Then the charge operator associated to $X$ is given by
\begin{equation}
\op{\mc{Q}}_{i^{0}}(X)\ket{\psi(y)}\defn -i\frac{d}{d\epsilon}\op{U}(\Lambda(\epsilon))\ket{\psi(y)}\bigg\vert_{\epsilon=0}
\end{equation}
By a straightforward calculation using \cref{eq:LorDI}, the action of the charge $\op{\mc{Q}}_{i^{0}}(X)$ is the sum of two terms 
\be
\label{eq:infLor}
    &\op{\mc Q}_{i^0}(X) \ket{\psi(y)} \\
     &\qquad = \sum_{n=0}^\infty~ \int_{\scri^n} d^3x_1 \cdots d^3x_n~ \bigg[ -i \oLie_X \psi^{A_{1}B_{1}\ldots A_{n}B_{n}}(x_1,\ldots,x_n;y) \op N_{A_{1}B_{1}}(x_1) \cdots \op N_{A_{n}B_{n}}(x_i) \ket{\omega_{\nu(y)}} \\
    &\qquad\quad + \psi^{A_{1}B_{1}\ldots A_{n}B_{n}}(x_1,\ldots,x_n;y) \op N_{A_{1}B_{1}}(x_1) \cdots \op N_{A_{n}B_{n}}(x_n) \op{\mc Q}_{i^0}(X'(y)) \ket{\omega_{\nu(y)}} \bigg]\, .
\ee
The first term arises from differentiating (with respect to $\epsilon$) the Lorentz transformation of the test tensors where $\mathfrak{L}$ is the Lie derivative acting on $\psi^{A_{1}B_{1}\ldots A_{n}B_{n}}(x_1,\ldots,x_n;y)$ viewed as a smooth function on $\scri^n \times \orbit_\Delta$. Explicitly, the generator $X$ can be decomposed as a vector field $Y$ on $\orbit_\Delta$ and a vector field $X^{A}$ on $\scri$ and so $\mathfrak{L}_{X}\psi$ can be expressed as
\be
    &\oLie_X \psi^{A_{1}B_{1}\ldots A_{n}B_{n}}(x_1,\ldots,x_n;y) \\
    &\qquad \qquad= \lb[ - \Lie_Y + \sum_{j=1}^n \lb( \tfrac{1}{2} u_j \ms D_C X_{j}^C \pd{u_j} + \Lie_{X_j} + \tfrac{1}{2} \ms D_C X_{j}^C \rb) \rb] \psi^{A_{1}B_{1}\ldots A_{n}B_{n}}(x_1,\ldots,x_n;y)
\ee
where each \(X_j^A\) is a copy of the vector field \(X^A\) on each of the \(\scri\) on which the test tensor depends. The second term in \cref{eq:infLor} arises from differentiating the Lorentz action on the vacua in \cref{eq:LorDI} with respect to $\epsilon$. Recall that $\Lambda'(y)$ is an element of the little group $\grp{L}_{\Delta(y)}$ at every point \(y \in \orbit_\Delta\) and therefore the action of the corresponding Lorentz charge is well-defined on each memory Fock space \(\Fock_{\Delta(y)}\). The Lorentz charge obtained by differentiating the little group unitary $\op U(\Lambda'(y))$ is $\op{\mc{Q}}_{i^{0}}(X'(y))$ where the Lorentz generator $X'(y)$ is given by (using \cref{eq:lorentz-little-projection})
\begin{equation}\label{eq:lorentz-little-gen-projection}
    X^{\prime}(y) \defn X + i (\pounds_{Y}\lambda(y))\lambda(y)^{-1}.
\end{equation}
Since states of the form of \cref{eq:nice-state} are dense in $\Fock_{\textrm{DI},\Delta}$ these Hilbert spaces are indeed representations of the full algebra $\ms{A}_{\Delta,\textrm{Q}}$. 

\subsection{Orbit spaces of memory under Lorentz transformations}
\label{sec:memory-orbits}

In the previous subsection, we gave a general construction of a large supply of states with memory that have well-defined BMS charges. These Hilbert spaces are obtained by a direct integral over the ``orbit space'' $\ms{O}$ of a chosen memory. In this section, we will enumerate the possible orbit spaces for any square-integrable memory. To specify the possible orbit spaces it will be convenient to choose complex stereographic coordinates $x^{A}=(z,\bar{z})$ on $\bb{S}^{2}$ which are related to the usual spherical coordinates by
\begin{equation}
z = e^{i\phi}\cot(\theta/2), \quad \quad \bar{z}=e^{-i\phi}\cot(\theta/2)
\end{equation}
Furthermore, it is also useful to introduce a complex null basis on the $2$-sphere
\begin{equation}
    m_{A} \equiv P^{-1}dz \eqsp \bar{m}_{A} \equiv P^{-1}d\bar z \eqsp q_{AB} = 2m_{(A} \bar m_{B)}
\end{equation}
which satisfy $m^{A}m_{A}=\bar{m}^{A}\bar{m}_{A}=0$ and $m^{A}\bar{m}_{A}=1$, and
\be\label{eq:P-defn-1}
    P(z,\bar z) \defn \frac{1 + z \bar z}{\sqrt{2}}.
\ee
In these coordinates, it is convenient to specify the action of the Lorentz group on \(\bb S^2\) by going to its double cover group \(\grp{SL}(2,\bb C)\) which consists of complex \(2 \times 2\) matrices of the form
\be
    \begin{pmatrix}a,&b \\ c,&d\end{pmatrix} \eqsp ad-bc =1 \eqsp z \mapsto \frac{az+b}{cz+d}
\ee
where the final relation is a M\"obius transformation of the coordinates \((z,\bar z)\). Note that both the identity matrix and its negative act trivially on \((z,\bar z)\), and the quotient of \(\grp{SL}(2,\bb C)\) by the central subgroup \(\bb Z_2\) containing these two elements is the Lorentz group \(\grp{S0}(1,3) \cong \grp{SL}(2,\bb C)/\bb Z_2\).

The memory tensor is symmetric and traceless and thus in this basis can be decomposed as 
\begin{equation}\label{eq:memory-basis}
    \Delta_{AB}(z,\bar z) = \Delta(z,\bar{z})m_{A}m_{B} + \bar{\Delta}(z,\bar{z}) \bar m_{A}\bar{m}_{B}
\end{equation}
in terms of a single, complex function $\Delta(z,\bar{z})$. Under a Lorentz transformation \(\Lambda \in \grp{SL}(2,\bb C)\) we have the transformation
\be\label{eq:memory-transformation}
    \Delta(z,\bar z) \mapsto (\Lambda\Delta)(z,\bar z) = \varpi_\Lambda e^{i 2 \phi_\Lambda} \Delta(\Lambda z, \bar{\Lambda z})
\ee
where the functions \(\varpi_\Lambda\) and \(\phi_\Lambda\) are given explicitly in \cref{eq:omega-phi}. The power of $\varpi_\Lambda$ and $e^{i  \phi_\Lambda}$ appearing in the Lorentz transformation \cref{eq:memory-transformation} refer to the fact that the complex function $\Delta$ has conformal weight \(w = -1\) and spin weight \(s = -2\) (see appendix \ref{sec:weighted-func} for the definition of these weights).

To find the little groups and orbits of the memory it is convenient to work with the function \(\Delta(z,\bar z)\) instead of the tensor field \(\Delta_{AB}\). Given a memory function $\Delta(z,\bar{z})$, the corresponding little group \(\grp{L}_\Delta\) is the maximal subgroup of $\grp{SL}(2,\bb C)$ which leaves $\Delta$ invariant according to the transformation law \cref{eq:memory-transformation}. Then the space of all other memory functions which can be obtained by acting on \(\Delta(z,\bar z)\) by all Lorentz transformations is parametrized by the orbit space \(\orbit_\Delta \cong \grp{SL}(2,\bb C)/\grp{L}_\Delta\). Furthermore, if $\grp{L}_{\Delta}$ is the little group of $\Delta$ then the little group of $\Lambda \Delta$ is the conjugation  $\Lambda \grp{L}_{\Delta}\Lambda^{-1}$. That is, conjugate little groups give rise to the same orbit space. So we need only to consider the conjugation classes of such little groups and with each orbit space $\ms{O}_{\Delta}$ and conjugacy class of $\grp{L}_{\Delta}$ is associated a representative ``invariant function'' $\Delta$ which is invariant under $\grp{L}_{\Delta}$.

We now summarize the classification of the possible orbit spaces of memory (the detailed analysis is collected in appendix \ref{sec:little-groups-orbit} and \ref{sec:orbit-stuff}). A trivial case is when the memory is invariant under the full Lorentz group $\grp{SL}(2,\bb{C})$. The only invariant function is \(\Delta = 0\) and the orbit space is just the one element set \(\set{0}\). In this case, the direct integral Hilbert space is the zero memory Fock space \(\Fock_0\).

Another simple case of physical interest is when the memory is non-vanishing but has no symmetries. This is the generic case that is expected to occur for any non-trivial scattering, e.g., if the in-state consists of an $n$-graviton state with vanishing memory then the out-state will generically have a memory that has no symmetries. A more precise argument to this effect is given in \cite{KP-BMS-particles}. In this case, the little group is the subgroup $\bb{Z}_{2}$ (consisting of the identity matrix and its negative) which has trivial action on $\bb{S}^{2}$. The orbit space is the identity component of the Lorentz group $\grp{SO}(3,1)$ which is homeomorphic to $\bb{R}^{3}\times \bb{R}\bb{P}^{3}$ where $\bb{R}\bb{P}^{3}$ is the sphere $\bb{S}^{3}$ with the antipodal points identified. Coordinates on this orbit space can be obtained directly from suitable coordinates on the Lorentz group (see \cref{eq:coord-defn}). These are Cartesian coordinates $(t,u,v)$ on $\bb{R}^{3}$ and standard spherical coordinates $(\theta,\phi,\psi)$ on $\bb S^{3}$. The Lorentz invariant measure on this orbit space is 
\begin{equation}
\label{eq:measureR3RP3}
    d\mu_{\bb{R}^{3}\times \bb{R}\bb{P}^{3}} = e^{2t}dt\, du\, dv\, d\phi\, d\psi\, d(\cos \theta).
\end{equation}

A less common case which may still be of some physical interest is when the memory is non-vanishing but invariant under a connected subgroup of $\grp{SL}(2,\bb{C})$ that is larger than $\bb{Z}_{2}$. Despite the fact there are many subgroups of $\grp{SL}(2,\bb{C})$, we prove in lemma \ref{lem:little-group-elimination} in appendix \ref{sec:little-groups-orbit} that in the gravitational case, where the spin weight of the memory function is non-zero, the only connected subgroup with a non-zero invariant function is the group $\Gamma$ which is the double cover of the group of rotations around a single axis. The invariant memory functions in this case are axisymmetric functions
\begin{equation}
    \Delta(z,\bar z) = e^{-i2 \arg(z)}\Delta(|z|)
\end{equation}
where the phase factor arises from the spin weight \(s = -2\) for the memory. The orbit space $\grp{SL}(2,\bb{C})/\Gamma \cong \bb{R}^{3}\times \bb{S}^{2}$ with invariant measure 
\begin{equation}
d\mu_{\bb{R}^{3}\times \bb{S}^{2}} = dt\, d\tilde u\, d\tilde v\, d\phi\, d(\cos \theta)
\end{equation}
where $(\theta,\phi)$ are spherical coordinates on $\bb{S}^{2}$ and \(\tilde u, \tilde v\) are new coordinates given explicitly in \cref{eq:tilde-u-v}. All other orbit spaces correspond to memory functions which are invariant under non-connected subgroups of $\grp{SL}(2,\bb{C})$. While such memories are not expected to generically arise in scattering theory, we classify these orbit spaces in appendix \ref{sec:little-groups-orbit}.

\begin{remark}[Generalization to massless scalars, QED and Yang-Mills]
\label{rem:qed-scalar-case}
The above analysis proceeds in an identical manner for the memories in the electromagnetic and massless scalar field cases. In the electromagnetic case the memory \(\Delta^\EM_A\) can be written in terms of a complex function (similar to \cref{eq:memory-basis}) with conformal weight \(w=-1\) and spin weight \(s = -1\). The only connected little groups with non-trivial memory functions are again $\Gamma$ and $\bb{Z}_{2}$ as in the gravitational case. For a massless scalar field, the memory function has conformal weight \(w=-1\) and spin weight \(s = 0\) and we additionally have the little group $\grp{SU}(2)$ leaving invariant any spherically symmetric memory $\Delta = \textrm{constant}$. The orbit space $\grp{SL}(2,\bb{C})/\grp{SU}(2) \cong \bb{R}^{3}$ is homeomorphic to the hyperboloid of future-directed unit-normalized timelike vectors in Minkowski spacetime with its standard, Lorentz invariant measure. The Yang-Mills memory $\Delta_{A,i}^{\YM}$, where ``\(i\)'' denotes an index in the Lie-algebra of the Yang-Mills group, is not invariant under either the Lorentz group or the group of large gauge transformations. Thus, the relevant orbit spaces correspond to quotients of this larger group. A more serious issue that arises in both Yang-Mills theories and massless QED is the production of ``collinear divergences'' which are not treated in this paper. 
\end{remark}

\begin{remark}[Direct integral over charges at spatial infinity in QED]
The analog of the supertranslation charges at spatial infinity are the so-called ``large gauge charges'' $\mc{Q}^{\EM}_{i^{0}}(x^{A})$ which is a function of conformal weight \(w=-2\) and spin weight \(s=0\) on the $2$-sphere. In the case of QED with massive charged fields, one can construct a Hilbert space $\Hilb_{\mc{Q}^{\EM}_{i^{0}}}$ of definite large gauge charge known as ``Faddeev-Kulish'' Hilbert spaces \cite{Kulish:1970ut}. However, while the Faddeev-Kulish Hilbert space with $\mc{Q}^{\EM}_{i^{0}} = 0$ has a unitary action of the Lorentz group, this is not the case for non-zero charge values since the charge is not Lorentz invariant. Nevertheless, we can follow a similar procedure as outlined in this paper to construct direct integral Hilbert spaces over the Lorentz orbit of the charge, which will have a unitary action of the Lorentz group. However, we do not see any good reason for the charges at spatial infinity to be square-integrable, so other types of orbit spaces can arise than the ones spelled out in this paper. Note that, the analogous Faddeev-Kulish Hilbert spaces do not exist in QED with massless charged fields or quantum gravity so this construction cannot be generalized to these cases \cite{PSW-IR}. 
\end{remark}

\acknowledgements
G.S. is supported by the Princeton Gravity Initiative at Princeton University. This work was supported in part by the NSF grant PHY-2107939 to the University of California, Santa Barbara. 

\appendix

\section{Little groups and orbits spaces of conformal-spin weighted functions on the $2$-sphere}
\label{sec:weighted-func}

In this appendix we collect the mathematical details of the conformal-spin weighted functions and the construction of their little groups and orbit spaces used in \cref{sec:memory-orbits} of the paper. The main arguments of the paper are concerned with the gravitational memory which can be represented by a square-integrable function $\Delta$ of conformal-spin weight \((w,s) = (-1,-2)\). However, the results in these appendices will apply to {\em any} square-integrable, conformal-spin weighted function with conformal weight \(w \leq -1\) and arbitrary spin weight. We will denote such a general, conformal-spin weighted function as $\varphi$. To begin we first review the basic structure of conformal-spin weighted functions and the regularity of the functions on $\bb{S}^{2}$. In appendix \ref{sec:little-groups-orbit}, we will then classify the associated connected little groups and orbit spaces of such functions. In appendix \ref{sec:orbit-stuff} we then construct cross-sections of the orbit spaces, its Lorentz generators and its invariant measure. Finally, in appendix \ref{sec:orbit-stuff} we consider the case of non-connected little groups and their corresponding orbit spaces. 

As in \cref{sec:memory-orbits}, we will use complex stereographic coordinates \((z,\bar z)\) on \(\bb S^2\). This coordinate patch covers the entire sphere except for one point which maps to \(z = \infty\). To cover the entire sphere we use a second coordinate patch \((z',\bar z')\) with the transition map \(z' = 1/z\) when \(z \neq 0,\infty\). In these coordinates the metric and area \(2\)-form on \(\bb S^2\) are given by
\be\label{eq:S2-metric}
    q_{AB} \equiv 2 P^{-2} dz d\bar z \eqsp \varepsilon_{AB} \equiv i P^{-2} d z \wedge d\bar z 
\ee
where
\be\label{eq:P-defn}
    P \defn \frac{1+ z \bar z}{\sqrt{2}}
\ee
We also choose a complex null basis give by
\be\label{eq:dyad-defn}
    m_A \equiv P^{-1} d z \eqsp \bar m_A \equiv P^{-1} d\bar z 
\ee
such that \(m^A m_A = \bar m^A \bar m_A = 0\), \(m^A \bar m_A = 1\), \(q_{AB} = 2 m_{(A} \bar m_{B)}\) and \(\varepsilon_{AB} = 2i m_{[A} \bar m _{B]}\).

We represent elements of the Lorentz group \(\grp{SL}(2,\bb C)\) by matrices as follows\footnote{The group \(\grp{SL}(2,\bb C)\) is the double cover of the identity-connected component of the Lorentz group \(\grp{SO}(1,3)\); we will refer to \(\grp{SL}(2,\bb C)\) as the Lorentz group for simplicity.}
\be\label{eq:matrix-form}
    \Lambda = \begin{pmatrix}a,& b \\ c,&d \end{pmatrix} \in \grp{SL}(2,\bb C) \eqsp a,b,c,d \in \bb C \eqsp ad-bc = 1
\ee
The action of any element \(\Lambda \in \grp{SL}(2,\bb C)\) on \(\bb S^2\) maps any point with coordinates \((z,\bar z)\) to another point \((\Lambda z, \bar{\Lambda z})\) according to\footnote{Compared to \cref{eq:z-transform}, the action of the Lorentz group used by McCarthy \cite{McCarthy1} is \(z \mapsto \frac{a z + c}{b z +d} = (J \cdot \Lambda^{-1} \cdot J^{-1} ) z\), where the \emph{inversion} \(J \in \grp{SL}(2,\bb C)\) is as defined in \cref{eq:inversion-defn}.}
\be\label{eq:z-transform}
    z \mapsto \Lambda z = \frac{az + b}{cz + d}
\ee
The transformations of the metric tensor and the null basis are obtained by pullback along the Lorentz transformation. An explicit computation using \cref{eq:S2-metric,eq:dyad-defn} gives
\begin{subequations}\label{eq:metric-dyad-transform}\begin{align}
    q_{AB}(z,\bar z) \mapsto (\Lambda q)_{AB}(z,\bar z) = q_{AB}(\Lambda z, \bar{\Lambda z}) &= \varpi_\Lambda^2(z,\bar z) q_{AB}(z,\bar z) \label{eq:q-transform} \\
    m_A(z,\bar z) \mapsto (\Lambda m)_A(z,\bar z) = m_A(\Lambda z, \bar{\Lambda z}) &= \varpi_\Lambda(z,\bar z) e^{i \phi_\Lambda(z,\bar z) } m_A(z,\bar z) \label{eq:m-transform}
\end{align}\end{subequations}
where
\begin{subequations}\label{eq:omega-phi}\begin{align}
    \varpi_\Lambda(z,\bar z) & = \frac{P(z,\bar z)}{P(\Lambda z, \bar{\Lambda z})} \lb(\td[(\Lambda z)]{z}\rb)^\half \lb(\td[(\bar{\Lambda z})]{\bar z}\rb)^\half = \frac{1+\abs{z}^2}{\abs{az+b}^2 + \abs{cz+d}^2} > 0 \label{eq:omega-h-defn} \\
    e^{i\phi_\Lambda(z,\bar z)} & = \lb(\td[(\Lambda z)]{z}\rb)^\half \lb(\td[(\bar{\Lambda z})]{\bar z}\rb)^{-\half} = \frac{\bar c \bar z +\bar d}{c z + d} \,. \label{eq:phi-h-defn}
\end{align}\end{subequations}
Henceforth, \(\varpi_\Lambda\) and \(\phi_\Lambda\) will always be understood as evaluated at the point \((z,\bar z)\), and we will drop this dependence from the notation.

Under the action of \(\Lambda \in \grp{SL}(2,\bb C)\) on \(\bb S^2\) any function \(\varphi(z,\bar z)\) transforms to another function \((\Lambda\varphi)(z,\bar z)\) by pullback as
\be
    \varphi(z,\bar z) \mapsto (\Lambda \varphi)(z,\bar z) = \varphi(\Lambda z, \bar{\Lambda z})
\ee
A function with conformal and spin weights depends not only on the point \((z,\bar z)\) but also on the choice of metric \(q_{AB}\) and the choice of basis \(m_A\); the weights specify how the function changes when the metric and basis are changed by a conformal transformation and a local rotation of the basis while keeping the points fixed, i.e.
\be
    q_{AB}(z,\bar z) \mapsto \varpi^2(z,\bar z) q_{AB}(z,\bar z) \eqsp m_A(z,\bar z) \mapsto \varpi(z,\bar z) e^{i \phi(z,\bar z)} m_A(z,\bar z)
\ee
A function \(\varphi[q,m](z,\bar z)\) has conformal weight \(w\) and spin weight \(s\) if
\be\label{eq:weights-defn}
    \varphi[\varpi^2 q, \varpi e^{i\phi} m](z,\bar z) = \varpi^w e^{is\phi} \varphi[q,m](z,\bar z)
\ee
where the metric and basis are also evaluated at the same point \((z,\bar z)\) as the function, and their indices have been suppressed. To compute the change of such a function under the transformation \(\Lambda \in \grp{SL}(2,\bb C)\) we also have to take into account the transformation of the metric and basis under the pullback map. We get
\be\label{eq:function-transform}
    \varphi[q, m](z,\bar z) \mapsto (\Lambda \varphi)[q,m](z,\bar z) &= \varphi[\varpi_\Lambda^{-2} q, \varpi_\Lambda^{-1}e^{-i\phi_\Lambda}m](\Lambda z,\bar{\Lambda z}) \\
    &= \varpi_\Lambda^{-w} e^{-is\phi_\Lambda} \varphi[q,m](\Lambda z,\bar{\Lambda z})
\ee
where we have used the fact that, from \cref{eq:metric-dyad-transform}, we have \(q_{AB}(z,\bar z) = \varpi_\Lambda^{-2} q_{AB} (\Lambda z, \bar{\Lambda z})\) and \(m_A(z,\bar z) = \varpi_\Lambda^{-1} e^{-i\phi_\Lambda} m_A(\Lambda z,\bar{\Lambda z})\) . Henceforth we do not write the explicit dependence of the function on the metric and basis for notational convenience, and denote the conformal-spin weights of a function by \((w,s)\).

The little groups and subsequent orbit spaces considered in \cref{sec:memory-orbits}, will strongly depend on the regularity of the conformal-spin weighted functions that we consider. As emphasized in the introduction (see the discussion around \cref{eq:memL2}), the relevant regularity is $L^{2}(\bb{S}^{2})$. We now consider how to specify the regularity of conformal-spin weighted functions on \(\bb S^2\) in terms of the stereographic coordinates. Since we need two coordinate patches, \((z,\bar z)\) and \((z',\bar z') = (1/z,1/\bar z)\), to cover \(\bb S^2\) we need to specify the regularity of the functions in both coordinate patches. This is conveniently done as follows: Consider the \emph{inversion} map
\be\label{eq:inversion-defn}
    J \defn \begin{pmatrix}0,&-1 \\ 1,& 0 \end{pmatrix} \in \grp{SL}(2,\bb C) \eqsp z \mapsto J z = - 1/z
\ee
The inversion of a function \(\varphi(z,\bar z)\) of weight \((w,s)\) is then given by (using \cref{eq:omega-phi,eq:function-transform})
\be\label{eq:function-inversion}
    (J \varphi)(z,\bar z) = (z/\bar z)^s \varphi(-1/z,-1/\bar z) 
\ee
The regularity of conformal-spin weighted functions on \(\bb S^2\) is specified by giving the regularity conditions on \(\varphi(z,\bar z)\) \emph{and} its inversion \((J \varphi)(z,\bar z)\) as functions on \(\bb C\). For instance, a function \(\varphi(z,\bar z)\) is smooth on \(\bb S^2\) if both \(\varphi(z,\bar z)\) and \((J \varphi)(z,\bar z)\) are smooth functions on \(\bb C\). Similarly, a function \(\varphi(z,\bar z)\) is in \(L^2(\bb S^2)\) (see remark \ref{rem:L2} below) if both \(\varphi(z,\bar z)\) and \((J \varphi)(z,\bar z)\) are \emph{locally} \(L^2\) in \(\bb C\), i.e. \(L^2\) over every bounded region in \(\bb C\), with respect to the natural measure on \(\bb S^2\) restricted to these coordinate patches. While the $L^{2}$ norm is not Lorentz invariant, the following remark illustrates that the $L^{2}$ topology on {\em any} conformal-spin weighted function is Lorentz invariant.  

\begin{remark}[\(L^2\) topology on weighted functions]
\label{rem:L2}
We remark that the \(L^2\) norm on conformally-weighted functions is not Lorentz invariant, except for the conformal weight \(w = -1\). However, for any function \(\varphi(z, \bar z)\) of conformal weight \(w\), consider the ``Lorentz transformed'' norm
\be
    \int_{\bb S^2} d\Omega(z, \bar z) \abs{(\Lambda \varphi)(z, \bar z)}^2 &= \int_{\bb S^2} d\Omega(z, \bar z)~ \varpi_\Lambda^{-2 w} \abs{\varphi(\Lambda z, \bar{\Lambda z})}^2 \\
    &= \int_{\bb S^2} d\Omega(\Lambda z, \bar{\Lambda z})~ \varpi_\Lambda^{-2(1+w)} \abs{\varphi(\Lambda z, \bar{\Lambda z})}^2
\ee
where \(d\Omega(z,\bar z)\) denotes the area element on \(\bb S^2\) at the point \((z,\bar z)\). Since \(\varpi_\Lambda(z, \bar z)\) is a \emph{strictly} positive smooth function on \(\bb S^2\), it is bounded both above and below by positive constants. Thus, for any \(\Lambda \in \grp{SL}(2,\bb C)\) and conformal weight \(w\), there exist positive constants \(c_{\Lambda,w}\) and \(C_{\Lambda,w}\) such that \(c_{\Lambda,w} \leq \varpi_\Lambda^{-2(1+w)}(z, \bar z) \leq C_{\Lambda,w}\). So we have
\be
    c_{\Lambda,w} \int d\Omega(z, \bar z) \abs{\varphi(z, \bar z)}^2 \leq \int d\Omega(z, \bar z) \abs{(\Lambda \varphi)(z, \bar z)}^2 \leq C_{\Lambda,w} \int d\Omega(z, \bar z) \abs{\varphi(z, \bar z)}^2
\ee
Thus, the ``Lorentz transformed'' norm is equivalent to the original \(L^2\) norm and the induced \(L^2\) topology is Lorentz invariant.
\end{remark}

\begin{remark}[Relation to the notation of Gel'fand, Graev and Vilenkin \cite{Gelfand5} and \(2\)-dimensional CFT]
\label{rem:Gelfand-CFT-notation}
Given a function \(\varphi(z, \bar z)\) of conformal-spin weights \((w,s)\), consider another function
\be
    \hat\varphi(z, \bar z) \defn P^{-w}(z, \bar z)\varphi(z, \bar z).
\ee
Using \cref{eq:P-defn,eq:omega-phi,eq:function-transform}, the transformation of \(\hat\varphi(z, \bar z)\) under the action of \(\Lambda \in \grp{SL}(2,\bb C)\) is given by
\be
    \hat\varphi(z, \bar z) \mapsto (\Lambda \hat\varphi)(z, \bar z) 
    & = (cz +d)^{w+s} (\bar c \bar z  + \bar d)^{w-s} \hat\varphi(\Lambda z, \bar{\Lambda z}).
\ee
Thus, in the notation of \cite{Gelfand5}, \(\hat\varphi(z, \bar z)\) is an irreducible representation of \(\grp{SL}(2,\bb C)\) denoted \(D_{(n_1,n_2)}\) with \((n_1,n_2) = (w+s+1,w-s+1)\). In the nomenclature used in \(2\)-dimensional CFT literature \(\hat\varphi(z,\bar z)\) would be a quasi-primary (or \(\grp{SL}(2,\bb C)\) primary) of \emph{conformal dimensions} \((h,\bar h)\) with \(-2h = n_1-1 = w+s \) and \(-2\bar h = n_2 -1 = w-s\) (see, e.g. \cite{Ginsparg}). Note that in the frequently used abuse of notation, \(\bar h\) is \emph{not} the complex conjugate of \(h\).

The inversion of \(\hat\varphi(z, \bar z)\) is then
\be
    (J \hat\varphi)(z, \bar z) = z^{w+s}~ {\bar z}^{w-s} \hat\varphi(-1/z, -1/\bar z) = z^{n_1-1}~ \bar z^{n_2-1} \hat\varphi(-1/z, -1/\bar z)
\ee
It can be then checked that the notion of smoothness of conformal-spin weighted functions described above is the same as the one in \cite{Gelfand5}.
\end{remark}

\subsection{Construction of the little groups and orbit spaces}
\label{sec:little-groups-orbit}

Given a function \(\varphi(z, \bar z)\) of conformal-spin weight \((w,s)\) we want to find its \emph{orbit} \(\orbit_\varphi\) under Lorentz transformations, i.e., \(\orbit_\varphi\) is the space of all functions \(\Lambda \varphi\) which can be obtained from \(\varphi\) by the action of some Lorentz transformation \(\Lambda\). Note that if there is some \(\Lambda' \in \grp{SL}(2,\bb C)\) which leaves \(\varphi\) invariant, i.e. \(\Lambda'\varphi = \varphi\), then \((\Lambda \Lambda') \varphi = \Lambda \varphi\). Thus, the transformations that leave \(\varphi\) invariant do not contribute new elements to its orbit space. So we define, the \emph{little group} \(\grp L_\varphi\) as the \emph{maximal} subgroup of \(\grp{SL}(2,\bb C)\) that leaves a given \(\varphi\) invariant. Thus the orbit space \(\orbit_\varphi\) is homeomorphic to the quotient of \(\grp{SL}(2,\bb C)\) by right multiplication by elements of the little group, i.e., \(\orbit_\varphi \cong \grp{SL}(2,\bb C)/\grp L_\varphi\). Further, if \(\grp L_\varphi\) is the little group of \(\varphi\), then the little group of \(\Lambda \varphi \in \orbit_\varphi\) is the conjugation \(\Lambda \grp L_\varphi  \Lambda^{-1}\). That is, conjugate little groups all give the same orbit space, and so we only care about the conjugacy classes of possible little groups.

The conformal-spin weights of the function \(\varphi\) and its regularity on \(\bb S^2\) play a crucial role in finding its little group. In the main paper the function \(\varphi\) we are interested in is the gravitational memory \(\Delta\) (see \cref{eq:memory-basis}) which has weights \((-1,-2)\). In remark \ref{rem:qed-scalar-case} we argued that the construction can also be generalized to the electromagnetic and massless scalar memories which have weights \((-1,-1)\) and \((-1,0)\), respectively. In each case, as explained in the introduction, the memory is required to be square-integrable on \(\bb S^2\) (see \cref{eq:memL2}), for the energy of the states to be well-defined. Thus, in the following, we will analyze the little groups and orbits of functions \(\varphi \in L^2(\bb S^2)\) with conformal weight \(w \leq - 1\) and arbitrary spin weight.\footnote{As noted in remark \ref{rem:qed-scalar-case}, one can also consider orbits of the electromagnetic charge in massive QED. Even though the conformal-spin weights for this case, \((w,s) = (-2,0)\), are covered in our general analysis, we do not know of any reason why the charges need to be square-integrable on the \(2\)-sphere, and thus there might arise more orbit types than analyzed here.}

In the main arguments of the paper, we are interested in constructing direct integrals over the orbit spaces once we have already obtained Hilbert spaces associated with some function \(\varphi\) (in our case, this function is the gravitational memory on null infinity). To do this, we will also need a measure on the orbit space which is at least quasi-invariant under Lorentz transformations, though in all cases of interest, it will be Lorentz-invariant. 

With the above considerations, we can organize the program into the following steps:
\begin{enumerate}
    \item Find all non-conjugate subgroups \(\grp L \subset \grp{SL}(2,\bb C)\). \label{step:subgroup}
    \item For each \(\grp L\), find a function \(\varphi(z, \bar z) \in L^2(\bb S^2)\) invariant under \(\grp L\) but not invariant under any larger subgroup containing \(\grp L\). \label{step:inv-func}
    \item \(\grp L = \grp L_\varphi\) is the little group of such a function \(\varphi(z, \bar z)\) and then its orbit space is \(\orbit_{\varphi} \cong \grp{SL}(2,\bb C)/\grp L_\varphi\). \label{step:orbit}
    \item Find a Lorentz-(quasi)-invariant measure on \(\orbit_\varphi\). \label{step:measure}
\end{enumerate}

For connected subgroups of \(\grp{SL}(2,\bb C)\), step \ref{step:subgroup} has already been completed (see \cite{BK, Kihlberg, Shaw}). The connected subgroups (up to conjugation) are given in Table~1 of \cite{McCarthy-nucl}. For these subgroups the inclusion into other subgroups is summarized in a ``lattice'' given in Table~3 of \cite{McCarthy-nucl} and reproduced in \cref{fig:subgroup-strc} below.\footnote{\label{fn:group-notation}The subgroup denoted by \(\Delta\) in \cref{fig:subgroup-strc} is isomorphic to \(\grp{E}(2)\), the isometry group of two-dimensional Euclidean space and the one denoted by \(\Lambda\) is isomorphic to \((\bb R,+)\) the group of real numbers under addition. We will use the latter notations so as to not confuse the little groups with our notation for the memory and Lorentz transformations.} As we will see, we will not need the specific enumeration of all groups appearing in \cref{fig:subgroup-strc} and, in the following, any specific subgroup considered will be explicitly defined. In this subsection we will work exclusively with the connected subgroups (and \(\bb Z_2\) since it has a trivial action on \(\bb S^2\)); the case of non-connected subgroups can also be analyzed following \cite{McCarthy2} and will be considered in appendix \ref{sec:non-connected}.

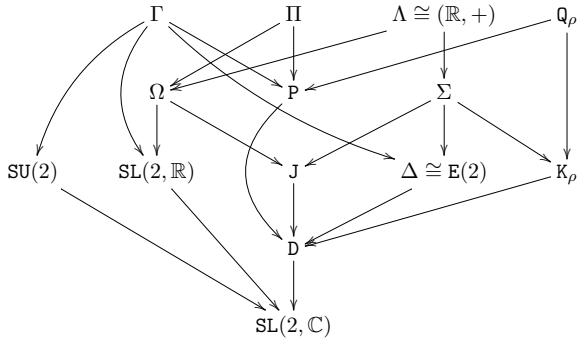
\begin{figure}[H]
\centerline{
\scalebox{0.75}{
\xymatrix{
	& \Gamma \ar@/_1pc/[ldd] \ar@/_1.5pc/[dd] \ar@/_1pc/[rrdd] \ar[rd] & \Pi \ar[ld] \ar[d] &\Lambda \cong (\bb R, +) \ar[lld] \ar[d] & \grp Q_\rho \ar[lld] \ar[dd] \\
	& \Omega \ar[d] \ar[rd] & \grp P \ar@/_2pc/[dd] & \Sigma \ar[ld] \ar[d] \ar[rd] \\
	\grp{SU}(2) \ar[rrdd] & \grp{SL}(2,\bb R) \ar[rdd] & \grp J \ar[d] & \Delta \cong \grp{E}(2) \ar[ld] & \grp K_\rho \ar[lld] \\
	&& \grp D \ar[d] \\
	&& \grp{SL}(2,\bb C)
}
}
}
\caption{Relations between the connected subgroups in the notation of \cite{McCarthy-nucl}. The arrows indicate inclusion (up to conjugation) as a subgroup, e.g. \(\Gamma \rightarrow \grp P\) indicates that \(\grp P\) contains a subgroup conjugate to \(\Gamma\). The center \(\bb Z_2\) and other disconnected subgroups have not been included. See also \cref{fn:group-notation}.}\label{fig:subgroup-strc}
\end{figure}

Proceeding to step \ref{step:inv-func}, we need to find a function \(\varphi(z, \bar z) \in L^2(\bb S^2)\) which is invariant under these subgroups. The following observations simplify the task greatly. The trivial case is of the function \(\varphi=0\) which has the little group \(\grp{SL}(2,\bb C)\). Next, since the action of \(\bb Z_2\) on \(\bb S^2\) is trivial, every function is \(\bb Z_2\)-invariant --- thus, any function which is not invariant under a bigger subgroup will have little group \(\bb Z_2\). This takes care of the most trivial and the most general cases. For the other cases, suppose \(\grp L_1 \subset \grp L_2 \subset \grp{SL}(2,\bb C)\) then any function invariant under \(\grp L_2\) must also be invariant under \(\grp L_1\). Contrapositively, if there is no non-zero function invariant under the subgroup \(\grp L_1\) then there is no non-zero function invariant under any subgroup (such as \(\grp L_2\)) containing \(\grp L_1\). With these facts in mind, we can employ the following strategy to find the functions invariant under some subgroup. We can start checking for the existence of non-zero invariant functions for the smallest subgroups (the topmost row in \cref{fig:subgroup-strc}) and work downwards. If no non-zero invariant square-integrable function is found for a particular subgroup, we can disregard all the groups in the diagram containing this subgroup until we reach the bottom, i.e., \(\grp{SL}(2,\bb C)\) whose invariant function is the zero function. As we will see below, this allows us to easily eliminate many of the subgroups in the diagram \cref{fig:subgroup-strc} and we are left with only a handful of cases that can arise as possible little groups.

The above strategy of ``weeding out'' the subgroups (from \cref{fig:subgroup-strc}) as possible little groups are summarized in the following lemma. This generalizes the analysis of McCarthy \cite{McCarthy1} from \(w = -3\) to all conformal weights \(w \leq -1\) and arbitrary spin weights.

\begin{lemma}\label{lem:little-group-elimination}
The only function \(\varphi \in L^2(\bb S^2)\), with conformal weight \(w \leq -1\) and arbitrary spin weight, which is invariant under the subgroups \(\Pi\), \((\bb R, +)\) and \(\grp Q_\rho\) is \(\varphi = 0\). 

\begin{proof}
    First, consider \(\Lambda = \lb(\begin{smallmatrix}e^{t(1+i\rho)/2} ,& 0 \\ 0,& e^{-t(1+i\rho)/2}\end{smallmatrix}\rb)\) for any \(t \in \bb R\). For a fixed real number \(\rho \neq 0\) it is an element of the one-dimensional subgroup \(\grp Q_\rho\) and for \(\rho = 0\) it is an element of the subgroup \(\Pi\). Any function which is invariant under the action of such an element \(\Lambda\) will satisfy
\be
    \varphi(z,\bar z) = \lb(\frac{1+ \abs{z}^2}{e^{-t} + e^t \abs{z}^2}\rb)^{-w} e^{- i s \rho t } \varphi(e^{t(1+i\rho)} z, e^{t(1-i\rho)} \bar z) \eqsp \forall~ t \in \bb R
\ee
Choose polar coordinates \(z = r e^{i\theta}\) and pick \(\Lambda\) such that \(e^t = 1/r\). Then, we have
\be
    \varphi(r,\theta) = \lb(\frac{1+ r^2}{2 r}\rb)^{-w} e^{is\rho \ln r} \varphi_0(\theta+\rho) \eqsp \varphi_0(\theta+\rho) \defn  \varphi(r = 1, \theta+\rho)
\ee
That is, \(\varphi\) is determined by scaling its value on the unit circle, except at \(z = 0, \infty\) which are the fixed points of this transformation. These fixed points are subsets of measure zero and so can be ignored\footnote{Note here that the regularity class of functions plays an important role in the analysis. For instance, subsets of measure zero cannot be ignored if one is allowing distributions instead of functions in \(L^2(\bb S^2)\).} for functions in \(L^2(\bb S^2)\). Next, we check whether such a function is in \(L^2(\bb S^2)\). Integrating over unit disk we get
\be
    \int_{\bb{S}^{2}}  d\Omega \abs{\varphi}^2 = \int_0^1 dr \int_0^{2\pi} d\theta~ \frac{r}{(1+r^2)^2} \lb(\frac{1+r^2}{2r}\rb)^{-2w} \abs{\varphi_0(\theta+\rho)}^2
\ee
This integral diverges near \(r = 0\) for \(w \leq -1\) unless \(\varphi_0(\theta+\rho) = 0\) almost everywhere. Thus, \(\varphi = 0\) as an element of \(L^2(\bb S^2)\).

Next, take the case \(\Lambda = \lb(\begin{smallmatrix}1,& i t \\ 0,& 1\end{smallmatrix}\rb) \in (\bb R, +)\) for any \(t \in \bb R\), so that any function invariant under \(\Lambda\) will satisfy
\be
    \varphi(z,\bar z) = \lb( \frac{1+\abs{z}^2}{\abs{z+it}^2 + 1} \rb)^{-w} \varphi(z+it, \bar z - it) \eqsp \forall~ t \in \bb R
\ee
Choose \(z = x + iy\) and pick \(\Lambda\) such that \(t = -y\) to get
\be
    \varphi(x, y) = \lb( \frac{1+ x^2 + y^2}{1+x^2} \rb)^{-w} \varphi_0(x) \eqsp \varphi_0(x) = \varphi(x, y = 0)
\ee
That is, the function is given by scaling its value on the real axis, except at the fixed point \(z = \infty\) which of measure zero. To check that this function is in \(L^2(\bb S^2)\) we integrate\footnote{Note, there is an error in the corresponding integral in \cite{McCarthy1} where the square of the conformal weight factor was not taken; however their final conclusion still holds.}
\be
    \int_{\bb{S}^{2}} d\Omega \abs{\varphi}^2 = \int_{\bb R^2} dx dy \frac{1}{(1+x^2+y^2)^2 }\lb( \frac{1+ x^2 + y^2}{1+x^2} \rb)^{-2w} \abs{\varphi_0(x)}^2
\ee
The \(x\)-integral can be made finite by choosing an appropriate \(\varphi_0(x)\) but the \(y\)-integral always diverges for \(w \leq -1\), unless \(\varphi_0(x) = 0\) almost everywhere. Thus, \(\varphi = 0\) as an element of \(L^2(\bb S^2)\).
\end{proof}
\end{lemma}

As a consequence of \ref{lem:little-group-elimination}, \(\Pi,(\bb R, +), \grp Q_\rho\) and \emph{all} subgroups of \(\grp{SL}(2,\bb C)\) containing these have the zero function as the invariant one. Then, following \cref{fig:subgroup-strc}, we only need to consider the cases where the little group is one of \(\bb Z_2\), \(\Gamma\), \(\grp{SU}(2)\) or \(\grp{SL}(2,\bb C)\). The case of \(\bb Z_2\) and \(\grp{SL}(2,\bb C)\) have already been discussed above. Thus, to complete steps \ref{step:inv-func} and \ref{step:orbit} it only remains to find square-integrable functions invariant under \(\Gamma\) and \(\grp{SU}(2)\), and their orbit spaces. This analysis is essentially the same as the one in \cite{McCarthy1}, which we detail next.

An element of \(\Gamma\) is given by the matrix \(\lb(\begin{smallmatrix} e^{i\phi/2},&0 \\ 0,& e^{-i\phi/2} \end{smallmatrix}\rb)\) so that the invariant function satisfies
\be
    \varphi(z,\bar z) = e^{-i s\phi} \varphi(e^{i\phi}z, e^{-i\phi} \bar z) \eqsp \forall~ \phi \in [0,4\pi)
\ee
Take \(z = r e^{i\theta}\) and choose \(\Lambda\) so that \(e^{i\phi} = e ^{-i\theta}\), so that
\be\label{eq:inv-Gamma}
    \varphi(r,\theta) = e^{is \theta} \varphi_0(r) \eqsp \varphi_0(r) = \varphi(r,\theta=0)
\ee
The condition that this function be in \(L^2(\bb S^2)\) gives
\be\label{eq:inv-Gamma-L2}
    2\pi \int_0^\infty dr \frac{r}{(1+r^2)^2} \abs{\varphi_0(r)}^2 < \infty
\ee
By choosing \(\varphi_0\) to satisfy suitable decay conditions as \(r \to \infty\) we can find many functions which satisfy this condition. It can also be verified that the inversion of such a function is also \(L^2(\bb S^2)\) by changing variables \(r \to 1/r\) in the above integral. So the functions invariant under \(\Gamma\) are of the form \(\varphi(z, \bar z) = e^{is\arg(z)} \varphi_0(\abs{z})\), where \(\varphi_0\) satisfies \cref{eq:inv-Gamma-L2}. The orbit of such a function under the action of Lorentz transformations is \(\grp{SL}(2,\bb C)/\Gamma\) which is homeomorphic to \(\bb R^3 \times \bb S^2 \).

Next we consider the subgroup \(\grp{SU}(2)\). Since \(\grp{SU}(2)\) acts transitively on \(\bb S^2\), any \(\grp{SU}(2)\)-invariant function is constant. However, since \(\Gamma \subset \grp{SU}(2)\) any \(\grp{SU}(2)\)-invariant function is also \(\Gamma\)-invariant of the form \cref{eq:inv-Gamma}. This implies that for \(s = 0\) the invariant function is \(\varphi = \text{constant}\) while for \(s \neq 0\) we must have \(\varphi = 0\). Thus, only for spin weight \(s = 0\) we have the little group \(\grp{SU}(2)\) and the orbit space \(\grp{SL}(2,\bb C)/\grp{SU}(2)\) which is homeomorphic to \(\bb R^3\).

For the little group \(\grp{SL}(2,\bb C)\) the invariant function is \(\varphi = 0\) and the orbit space is the one element set \(\set{0}\). Finally as discussed above, any function which cannot be transformed into any of the above functions by a Lorentz transformation must have \(\bb Z_2\) as the largest subgroup leaving it invariant, since \(\bb Z_2\) just acts trivially. In this case the orbit space is the identity component of the Lorentz group \(\grp{S0}(1,3)\) which is homeomorphic to \(\bb R^3 \times \bb{RP}^3\), where \(\bb{RP}^3\) is the real projective space obtained by identifying the antipodal points on \(\bb S^3\).

\subsection{Cross-sections, Lorentz generators and invariant measures on orbits}
\label{sec:orbit-stuff}

Next, given a little group \(\grp{L}\) and the corresponding orbit \(\orbit\), we outline how to compute the cross-sections \(\lambda : \orbit \to \grp{SL}(2,\bb C)\), the vector fields on \(\orbit\) generating the action of the Lorentz group on \(\orbit\) and the Lorentz-invariant measures on \(\orbit\). This choice of smooth cross-section together with the vector fields on \(\orbit\) as well as its invariant measure were crucially used in \cref{subsec:gencon} to explicitly obtain a strongly continuous action of the Lorentz group on $\Fock_{\textrm{DI},\Delta}$. In the following, we will closely follow the treatment given in \cite{BK, Kihlberg}, with some changes to the notation.

The generators of the Lie algebra \(\mf{sl}(2, \bb C)\) can be written in terms of the Pauli matrices as \(L_{ij} = \half \epsilon_{ij}{}^k\sigma_k\) and \(L_{0i} = \nfrac{i}{2} \sigma_i\) for \(i,j,k = 1,2,3\). Choose coordinates \(x^\mu = (t,u,v,\phi,\psi,\theta)\) on \(\grp{SL}(2,\bb C)\), so that any Lorentz transformation \(\Lambda\) can be written as (note, the Einstein summation convention is not used in writing \(x^\mu L_\mu\))
\be\label{eq:coord-defn}
    \Lambda(x^\mu) = \prod_{x^\mu} e^{-i x^\mu L_\mu} \defn e^{-i \phi L_{12}} e^{-i\theta L_{31}} e^{-i\psi L_{12}} e^{i t L_{03}} e^{-i u \lb( L_{01}+L_{31} \rb)} e^{-i v \lb( L_{02}-L_{23} \rb)} \,.
\ee
where \((\phi,\theta,\psi)\) are Euler angles parametrizing rotations in \(\grp{SU}(2)\) and \((t,u,v)\) parametrize the Lorentz boosts. The ranges of the coordinates given by
\be\label{eq:coord-range}
    -\infty < t,u,v < \infty \eqsp 0 \leq \theta \leq \pi \eqsp 0 \leq \phi + \psi < 4\pi \eqsp -2\pi \leq \phi - \psi < 2\pi
\ee
If instead \(\Lambda\) is in the identity component of \(\grp{S0}(1,3)\) we have the restricted range for the coordinates \(0 \leq \phi,\psi < 2\pi\). In terms of these coordinates the \(2\times 2\) matrix form (used previously) is given by\footnote{Note that we have corrected a sign error in the second entry in the top row from \cite{BK}.}
\be
    \Lambda(x^\mu) =
    \begin{pmatrix}
    e^{-t/2} e^{-i (\phi+\psi)/2 } \cos\nfrac{\theta}{2} - e^{t/2} e^{-i(\phi-\psi)/2}(u+iv) \sin\nfrac{\theta}{2},& - e^{t/2} e^{-i(\phi-\psi)/2} \sin\nfrac{\theta}{2} \\
    e^{-t/2} e^{i (\phi-\psi)/2 } \sin\nfrac{\theta}{2} + e^{t/2} e^{i(\phi+\psi)/2}(u+iv) \cos\nfrac{\theta}{2},& - e^{t/2} e^{i(\phi+\psi)/2} \cos\nfrac{\theta}{2}
    \end{pmatrix}
\ee

The coordinates \(x^\mu\) induce coordinates \(y^\mu\) on each orbit space corresponding to a little group \(\grp{L}\) as follows. Recall that the orbit space is obtained by the quotient \(\orbit = \grp{SL}(2,\bb C)/\grp{L}\), i.e., the orbit consists of the equivalence classes under right multiplication by elements of \(\grp{L}\). Let \({x'}^\mu\) be the subset of the coordinates \(x^\mu\) which parametrize the elements of \(\grp{L}\), and \(y^\mu\) be the remaining ones. Using the commutation relations of the generators we rewrite \cref{eq:coord-defn} so that the generators corresponding to the little group \(\grp{L}\) are all to the right, i.e.
\be\label{eq:coord-defn-little}
    \Lambda(x^\mu) = \prod_{{y^\mu}} e^{-i y^\mu L_\mu} \prod_{{x'}^\mu} e^{-i {x'}^\mu L_\mu} \eqsp e^{-i {x'}^\mu L_\mu} \in \grp{L}
\ee
Now note that right multiplication by any element of \(\grp{L}\) only changes the parameters \({x'}^\mu\) and thus the coordinates \(y^\mu\) parametrize the orbit \(\orbit\). The point \(y_0 \equiv y^\mu = 0\) corresponds to \(\prod\limits_{{y^\mu}} e^{-i y^\mu L_\mu} \) being the identity i.e., it labels the equivalence classes of Lorentz transformations up to right multiplication by the little group \(\grp L\).

With these coordinates on \(\orbit\) one gets an obvious choice for the smooth cross-section \(\lambda : \orbit \to \grp{SL}(2,\bb C)\) as
\be
    \lambda(y) = \Lambda(x^\mu)\big\vert_{{x'}^\mu = 0} = \prod_{{y^\mu}} e^{-i y^\mu L_\mu}
\ee
and from \cref{eq:coord-defn-little} we see that \(\lambda(y)y_0 = y\) as required.

Any element of the Lie algebra \(\mf{sl}(2,\bb C)\) can be represented by a vector field on \(\grp{SL}(2,\bb C)\). The explicit form for the various generators as vector fields \(X^\mu \pd{x^\mu}\) in the coordinates \(x^\mu\) of \cref{eq:coord-defn} can be found in \cite{BK}. In the coordinates adapted to the little group \(\grp{L}\) we have
\be
    X^\mu \pd{x^\mu} = Y^\mu \pd{y^\mu} + {X'}^\mu \pd{{x'}^\mu} \in \mf{sl}(2,\bb C)
\ee
The components \(Y^\mu \pd{y^\mu}\) define the vector fields on the orbit which are projections of the generators of the Lorentz group.

To find the invariant measure on the orbit we can proceed as follows. The invariant Haar measure on the group \(\grp{SL}(2,\bb C)\) is given by (see \cite{Gelfand5,BK})
\be\label{eq:full-measure}
    d\mu(\Lambda) = \abs{a}^{-2} d^2a\, d^2b\, d^2c = e^{2t} dt\, du\, dv\, d\phi\, d\psi\, d(\cos\theta)
\ee
where the first formula is given in terms of the matrix parameterization \cref{eq:matrix-form} (with \(d^2a\) being the Lebesgue measure on \(\bb C\)) and the second equality is expressed in terms of the coordinates defined in \cref{eq:coord-defn}. To find the measure on the orbits we again rewrite the group element \(\Lambda\) as in \cref{eq:coord-defn-little} with the generators of the little group to the right, and express the second form in \cref{eq:full-measure} as \(d\mu(\Lambda) = d\mu_\orbit(y) d\mu(x')\) and the \(d\mu_\orbit(y)\) is the invariant measure on the orbit \(\orbit\) (see \cite{McCarthy-nucl,BK}).

Let us look at some example computations in the procedure described above. Consider the little group \(\grp{L} = \Gamma\) whose elements are of the form \(e^{-i\psi L_{12}}\). To find the orbit we move this generator to the right in expression in \cref{eq:coord-defn}. Using the commutation relations of the Lorentz algebra we get
\be
    \Lambda = e^{-i \phi L_{12}} e^{-i\theta L_{31}}  e^{-i \tilde u \lb( L_{01}+L_{31} \rb)} e^{-i \tilde v \lb( L_{02}-L_{23} \rb)} e^{i t L_{03}} e^{-i\psi L_{12}}
\ee
where the new coordinates \((\tilde u, \tilde v)\) are given  by
\be\label{eq:tilde-u-v}
    \begin{pmatrix} \tilde u \\ \tilde v \end{pmatrix} =
    e^t\begin{pmatrix}
    \cos\psi, & - \sin\psi \\
    \sin\psi, & \cos\psi
    \end{pmatrix}
    \begin{pmatrix} u \\ v \end{pmatrix}
\ee
Then, the coordinates on the orbit space \(\orbit \cong \grp{SL}(2,\bb C)/\Gamma\) are given by \(y^\mu = (t,\tilde u, \tilde v,\phi,\theta)\). This immediately tells us that the orbit is homeomorphic to \(\bb R^3 \times \bb S^2\) and the measure on this orbit is \(dt\, d\tilde u\, d\tilde v\, d\phi\, d(\cos\theta)\).

Similarly, for the little group \(\grp{SU}(2)\) it is convenient to instead write
\be
    \prod_{{y^\mu}} e^{-i y^\mu L_\mu} =
    \begin{pmatrix}
    k_0 + k_3 ,& k_1 + i k_2 \\
    k_1 - i k_2 ,& k_0 - k_3
    \end{pmatrix} \eqsp
    \text{ with } -(k_0)^2 + (k_1)^2 + (k_2)^2 + (k_3)^2 = -1
\ee
so that \cref{eq:coord-defn-little} corresponds to a polar decomposition of the \(\grp{SL}(2,\bb C)\) matrix. Thus, in this case, the orbit can be identified with the hyperboloid of future-directed unit-normalized timelike vectors in Minkowski spacetime and the measure is the natural Lorentz-invariant measure on this hyperboloid.

We collect the little groups, the invariant functions, the homeomorphism type of the orbit space and the invariant measure in table \ref{tab:orbits-table}. In each case, the invariant function is of the form indicated in the second column and cannot be transformed by a Lorentz transformation into the form given in any of the previous rows.

\begin{table}[!h]
\centering
\begin{tabular}{|c|c|c|c|}
    \hline
    Little group & Invariant function & Orbit homeomorphic to & Invariant measure \\
	\hline\hline
    \(\grp{SL}(2,\bb C)\) & \(\varphi = 0\) & \(\set{0}\) & \\\hline
    \(\grp{SU}(2)\) & \(\varphi = \text{constant}\), \(s = 0\) & \(\bb R^3\) & \( d^3\vec k/\sqrt{1+\vec k^2} \) \\\hline
    \(\Gamma\) & \(\varphi = e^{is \arg(z)} \varphi_0(\abs{z})\) & \(\bb R^3 \times \bb S^2\) & \(dt\, d\tilde u\, d\tilde v\, d\phi\, d(\cos\theta)\) \\\hline
    \(\bb Z_2\) & \(\varphi\) & \(\bb R^3 \times \bb{RP}^3\) & \(e^{2t} dt\, du\, dv\, d\phi\, d\psi\, d(\cos\theta)\) \\
    \hline
\end{tabular}
\caption{Summary of the (connected) little groups, invariant functions, homeomorphism class of the orbit and invariant measures for a function of conformal-spin weight \((w,s)\).}
\label{tab:orbits-table}
\end{table}

\subsection{Non-connected little groups}
\label{sec:non-connected}

We briefly remark on the non-connected subgroups of \(\grp{SL}(2,\bb C)\) that arise as possible little groups. The case of the center \(\bb Z_2\) has already been considered since it acts trivially on \(\bb S^2\). Following, McCarthy \cite{McCarthy2} it can be shown that the only other non-connected groups which arise as possible little groups are the lifts to \(\grp{SU}(2)\) of non-connected subgroups of the rotation group \(\grp{SO}(3)\). These are considered below.

First, we consider the group \(\Theta\), which consists of elements of \(\Gamma\) along with all their group products with the inversion \(J\) (see \cref{eq:inversion-defn}). Every element of \(\Theta\) is of the form
\be
    R(\psi) = \begin{pmatrix} e^{i\psi/2},& 0 \\ 0,& e^{-i\psi/2} \end{pmatrix} \eqsp J(\psi) = \begin{pmatrix} 0,& -e^{-i\psi/2} \\ e^{i\psi/2}, & 0 \end{pmatrix} = J R(\psi)
    \eqsp \text{ for } \psi \in [0,4\pi)
\ee
where \(R(\psi) \in \Gamma\). Thus the functions invariant under \(\Theta\) are those that are invariant under both \(\Gamma\) and the inversion \(J\). The \(\Gamma\)-invariant function has already been found to be of the form
\be
    \varphi(z,\bar z) = e^{is \arg(z)} \varphi_0(\abs{z})
\ee
Now we impose invariance under the inversion \(J\) which gives the condition
\be
    \varphi(z,\bar z) &= e^{i2s \arg(z)} \varphi(-1/z, -1/\bar z) \\
    \varphi_0(\abs{z}) &= (-1)^s \varphi_0(1/\abs{z}) \\
\ee
where in the second line we have substituted the \(\Gamma\)-invariant form of the function. Now repeated application of the relation above gives
\be
    \varphi_0(\abs{z}) = (-1)^s \varphi_0(1/\abs{z}) = (-1)^{2s} \varphi_0(\abs{z}) \text{ for all } z \in \bb C
\ee
which is consistent only for integer spin weight, \(s \in \bb Z\). So the \(\Theta\)-invariant function has \(s \in \bb Z\) and is of the form
\be
    \varphi(z,\bar z) &= 
    \begin{cases}
    e^{is \arg{z}} \varphi_0(\abs{z}) ,& \abs{z} < 1 \\
    (-1)^s e^{is \arg{z}} \varphi_0(1/\abs{z}) ,& \abs{z} > 1
    \end{cases}
\ee
This function has possible discontinuities on \(z \in [0,1] \union [-1,-\infty] \union \set{\abs{z} = 1}\) which is a set of measure zero and can be verified to be in \(L^2(\bb S^2)\) with suitable conditions on \(\varphi_0(\abs{z})\).

The next non-connected group that arises is the \emph{cyclic group} \(C_n \cong \bb Z_n \subset \Gamma\), for a fixed integer \(n \geq 1\), which has elements of the form \(\lb(\begin{smallmatrix}e^{i2\pi/n},&0 \\ 0,& e^{-i2\pi/n} \end{smallmatrix}\rb)\). This is the lift to \(\grp{SU}(2)\) from \(\grp{SO}(3)\) of the group of all rotation symmetries of a regular \(n\)-sided polygon. The analysis here is similar to the case of \(\Gamma\), and the invariant functions are functions periodic under rotations with period \(4\pi/n\):
\be
    \varphi(z,\bar z) = e^{i s \arg(z)} \varphi_0(z,\bar z) \text{ such that } \varphi_0(z, \bar z) = \varphi_0(e^{i 4\pi/n}z , e^{-i4\pi/n} \bar z)
\ee

The next one is the \emph{binary dihedral group} \(\tilde D_n\), for a fixed integer \(n \geq 1\), which consists of elements of \(C_n\) along with all their group products with the inversion \(J\). This is the lift to \(\grp{SU}(2)\) from \(\grp{SO}(3)\) of the group of all rotation and reflection symmetries of a regular \(n\)-sided polygon. The analysis is the same as in the case of \(\Theta\) considered above with \(\Gamma\) replaced by the subgroup \(C_n\). The invariant function is
\be
    \varphi(z,\bar z) &= 
    \begin{cases}
    e^{is \arg{z}} \varphi_0(z,\bar z) ,& \abs{z} < 1 \\
    (-1)^s e^{is \arg{z}} \varphi_0(-1/z,-1/\bar z) ,& \abs{z} > 1
    \end{cases}
\ee
where \(s \in \bb Z\) and \(\varphi_0(z, \bar z) = \varphi_0(e^{i 4\pi/n}z , e^{-i4\pi/n} \bar z)\).

The remaining three non-connected subgroups are the binary tetrehedral group, the binary octahedral group and the binary icosahedral group, which are lifts to \(\grp{SU}(2)\) from \(\grp{SO}(3)\) of the symmetry groups of a regular tetrahedron, a regular octahedron (or a cube) and a regular icosahedron (or a dodecahedron), respectively. For spin weight \(s = 0\) these groups arise as little groups (see \cite{McCarthy2}), however we have not been able to show if these groups arise as little groups for any non-zero spin weighted functions.

For these groups the orbit spaces can be found by taking the quotient of the ones already found by the group actions and can be found in Table.~2 of \cite{McCarthy2}.



\bibliographystyle{JHEP}
\bibliography{asymp-quantization}      
\end{document}